\documentclass[twocolumn]{emulateapj}
\usepackage{graphicx}
\newcommand{\be}{\begin{equation}}
\newcommand{\ee}{\end{equation} }
\newcommand{\ba}{\begin{eqnarray}}
\newcommand{\ea}{\end{eqnarray}}

\newcommand{\bbeta}{\mbox{\boldmath$\beta$}}

\newcommand{\nn}{\mbox{} \nonumber \\ \mbox{} }
\newcommand{\kB}{k_{\rm B}}

\shorttitle{Tiny Electromagnetic Explosions}
\shortauthors{Thompson}	
\begin{document}
\title{Tiny Electromagnetic Explosions}
\author{Christopher Thompson}
\affil{Canadian Institute for Theoretical Astrophysics, 60 St. George St., Toronto, ON M5S 3H8, Canada.}		

\begin{abstract}
This paper considers electromagnetic transients of a modest total energy (${\cal E} \gtrsim 10^{40}$ erg) 
and small initial size (${\cal R} \gtrsim 10^{-1}$ cm).  They could be produced during collisions
between relativistic field structures (e.g. macroscopic magnetic dipoles)
that formed around, or before, cosmic electroweak symmetry breaking.
The outflowing energy has a dominant electromagnetic component; a subdominant thermal component (temperature $> 1$ GeV)
supplies inertia in the form of residual $e^\pm$.   A thin shell forms that expands subluminally, attaining a
Lorentz factor $\sim 10^{6-7}$ before decelerating.  Drag is supplied by the reflection of an ambient magnetic
field, and by deflection of ambient free electrons.  Emission of low-frequency (GHz-THz) superluminal waves takes
place through three channels:  i) reflection of the ambient magnetic field;  ii)  direct linear conversion of
the embedded magnetic field into a superluminal mode; and iii) excitation outside the
shell by corrugation of its surface.  The escaping electromagnetic pulse is very narrow (a few wavelengths)
and so the width of the detected transient is dominated by propagation effects.  
GHz radio transients are emitted from i) the dark matter halos of galaxies
and ii) the near-horizon regions of supermassive black holes that formed by direct gas collapse and
now accrete slowly.  Brighter and much narrower 0.01-1 THz pulses are predicted at a rate at least comparable
to fast radio bursts, experiencing weaker scattering and absorption.  The same explosions also accelerate protons 
up to $\sim 10^{19}$ eV and heavier nuclei up to $10^{20-21}$ eV.
\end{abstract}

\section{Introduction}

Consider the release of a large energy in electromagnetic fields in a small volume.
The explosion that results may be described as {\it tiny} if 
\vskip .05in
\noindent 1. The initial impulse is narrower than the wavelength of the radiation that is eventually 
detected by the observer, meaning that energy is transported outward in a very thin shell; and 
\vskip .05in
\noindent 2. Ambient charged particles are deflected, but not fully reflected, by the magnetic field embedded
in this expanding shell.
\vskip .05in

The motivation for this study comes from the detection of bright millisecond-duration radio transients 
\citep{lorimer07,thornton13}, which appear to originate at cosmological
distances \citep{ravi16,chatterjee17,marcote17,tendulkar17}, and repeat in
at least one case \citep{spitler16,scholz16}.  The implied energy in GHz frequency radiation, if emitted isotropically, 
can approach $10^{40}$ erg.
Although many astrophysical scenarios have been proposed to account for FRBs, none is supported by ab initio
emission calculations.   This phenomenon may simply represent the extreme tail of the giant pulse process
observed in high-voltage radio pulsars \citep{cordes16}, but if FRBs are truly of cosmological origin this involves an
extrapolation of several orders of magnitude in pulse energy.  

Brief ($\sim $ millisecond) transients of a much greater energy (exceeding $10^{46}$ erg) are detected 
from Galactic magnetars \citep{wt06,tzw15}, but appear to be powered by relatively long-wavelength magnetic disturbances,
some $10^{5-6}$ times larger than the radio wavelengths at which FRBs are detected.
The nanosecond duration pulses detected from high-voltage radio pulsars have a brightness temperature that
competes with those inferred for FRBs -- but a much lower energy, and an emitting volume not be much larger than 
a radio wavelength (e.g. \citealt{hankins03}).  Merging binary neutron
stars and collapsing magnetars also release enormous energy in large-scale electromagnetic fields, but the
brightness of the associated radio emission is strongly limited by induced Compton scattering off ambient
plasma.  

We therefore must take seriously the possibility that FRBs represent a new type of physical phenomenon,
representing the decay or annihilation of
objects smaller than a radio wavelength.   From a total energy $\gtrsim 10^{40}$ erg and size $< 10$ cm 
one immediately infers energy densities even higher than the rest energy density of macroscopic nuclear 
matter, which itself is much too `dirty' to act as a plausible source of FRBs.  

The simplest stable field structure that could store this energy is a magnetic field.  Flux densities
around or above $\sim 10^{20}$ G are now required, well in excess of those present in magnetars 
($\sim 10^{15}$ G).  Although the spontaneous decay of macroscopic magnetized objects could be considered, one 
is quickly led to examine collisions, as mediated by the intermediate-range dipole-dipole interaction.
The collision speed approaches the speed of light if most of the dipole mass is
electromagnetic in origin.  Some form of superconductivity is required for the supporting electric current
to persist over cosmological timescales.  Hence we adopt the acronym `Large Superconducting Dipole' 
(LSD for short: \citealt{t17}, hereafter Paper II).

Macroscopic magnetic dipoles could arise from symmetry breaking at energy scales ranging from $\sim 1$ TeV up to
the grand unification scale ($10^{15}$-$10^{16}$ GeV).   An old suggestion involves
static loops of current-carrying cosmic string \citep{witten85}, which if formed early in the cosmic expansion
could come to dominate the mass density \citep{otw86,copeland87,haws88,davis89}.   Within these cosmic `springs' the string
tension force is counterbalanced by a positive pressure arising from the magnetic field and the kinetic energy of the
charge carriers.  A GUT-scale `spring' loop of radius ${\cal R} \sim 10^{-1}$-1 cm and 
mass per unit length $\mu_s \sim 10^{20}$-$10^{22}$ g cm$^{-1}$ (corresponding to $G\mu_s/c^2 \sim 10^{-8}$-$10^{-6}$) 
would support a magnetic field $10^{20}$-$10^{22}$ G at a distance $\sim {\cal R}$ from the string.  This is many orders
weaker than the field in the string core, and weak enough to qualify as an ordinary magnetic field 
(as opposed e.g. to a hypermagnetic field).  
Stable dipoles arising from a symmetry breaking process near the electroweak scale would have a 
similar relation between size and mass, although we are not aware of concrete models describing them.
The magnetic field within such a structure ($\sim 10^{26}$ G) could actually exceed the macroscopic 
field around a GUT-scale `spring' loop, allowing a somewhat smaller size.   

Emission at a wavelength $\lambda \gg {\cal R}$ is a natural consequence of the radial spreading of the energy
pulse produced by the partial annihilation of two dipoles, although at the cost of a reduction
in electromagnetic fluence by a factor $\sim ({\cal R}/\lambda)^{1/2}$.
This paper is devoted to understanding the dynamics and radiative properties of such tiny electromagnetic
explosions, and the astrophysical environments in which they may occur.  Extremely relativistic motion
is involved, and the dissipation is spread out over dimensions ranging between $\sim 10^8$ and $10^{12}$ cm, depending
on the ambient magnetic pressure and free electron density.

Previous theoretical work devoted to exploding black holes is partly relevant here.  
Although the energy released is much too small to explain cosmological FRBs, the expanding cloud of charged
relativistic particles produced will have a similar interaction with an ambient magnetic field \citep{rees77,blandford77}.
Attempts have been made to connect FRBs with evaporating black holes, but the overall energy scale is wrong
without substantial modifications of gravity \citep{barrau14}.  On the other hand, electromagnetic pulses
from cosmic string cusps forming on loops of size $L$ persist for a large number 
$\sim (L/\lambda)^{2/3}$ of wave periods \citep{vachas08,yu14}.  The energy deposition is then far from 
`tiny', and the approach to radio frequency emission taken in this paper is not applicable.  In addition, keeping 
the pulse width shorter than $\sim 1$ ms (about $10^6$ wave periods in the GHz band) requires the
an emitting loop of GUT-scale string to be substantially smaller than the mean size as determined by gravitational 
radiation energy loss.

The energy pulse produced by a collision between LSDs will be partitioned between 
electromagnetic fields of coherence length $\sim {\cal R}$ and thermal plasma.  
As we show here, various emission channels for GHz-THz radiation are available if most of the energy released
is electromagnetic:  in addition to reflection of the ambient magnetic field, there is a possibility of 
direct linear conversion of the ejected magnetic field to a superluminal electromagnetic mode, and also 
excitation of such a mode by a corrugation of the shell surface.

The low total energy released by GHz FRBs ($\sim 10^{-18}$ of the energy density in the dark matter) 
is easily explained by the rarity of collisions between LSDs, without invoking extremely long
lifetimes or low abundances.  
Important differences also arise in the expected environments of exploding primordial black holes and
colliding LSDs.  The black holes, if present, would trace the dark matter distribution within
galaxies.   Dipoles with the mass and size inferred here would experience only weak drag off the
interstellar medium (ISM), but a small fraction would be trapped in a high-density cusp surrounding
supermassive black holes (SMBHs) that form by the direct collapse of dense gas clouds (Paper II).
This then opens up a mechanism for producing both repeating and singular FRB sources, with the repeating
sources potentially dominating the overall rate and concentrated closer to the epoch of SMBH formation

Near its emission zone the electromagnetic pulse is much narrower than the observed FRBs.  In Paper II, we show
how the pulse can be significantly broadened by interaction with very dense plasma near the emission site, as well
as by more standard multi-path propagation through intervening plasma.   
GHz radiation produced by collisions is only detectable if the ambient plasma density is low enough to allow transparency.  
If repeating FRBs do indeed
originate from such dense zones as the near-horizon regions of SMBHs, then there is the interesting suggestion
of much narrower, and even brighter, pulses in the 0.01-1 THz range that reach the Earth at an even higher rate
than the GHz FRBs.

The plan of this paper is as follows.  The prompt dynamics of an initially tiny
relativistic shell composed of thermal plasma and a strong magnetic field
is examined in Section \ref{s:explosion}.  The interaction of this shell with ambient cold plasma is
considered in some detail in Section \ref{s:decel}.  Various channels for the emission of a low-frequency
electromagnetic wave are described in Section \ref{s:radio}, and the modification of the pulse by propagation
is examined in Section \ref{s:prop}.  The following Sections \ref{s:highfreq0} and \ref{s:uhecr} deal with 
the emission of higher-frequency radio waves, gamma-rays, and ultra-high energy cosmic ray ions.
Our conclusions and predictions are summarized in Section \ref{s:summary}.  The Appendix gives further details
of the interaction of ambient charged particles with a strong electromagnetic pulse.

\section{Shell Acceleration}\label{s:explosion}

Consider the release of energy ${\cal E}$ in electromagnetic fields and relativistic particles within a small 
radius ${\cal R}$.  The outgoing pulse is approximately spherical with thickness $\Delta R_s \sim {\cal R}$
and duration $\sim {\cal R}/c$.   Here we normalize ${\cal E}$ to $10^{41}$ erg, corresponding to a
rest mass $10^{20}$ g.  The choice of ${\cal R}$ is informed by the requirement that the initial energy
density correspond to some physical scale $E_{\rm LSD}$.  If this lies above the electroweak scale,
then ${\cal R}$ is smaller than a radio wavelength,\footnote{Throughout this paper we use the shorthand 
$X = X_n \times 10^n$, where quantity $X$ is expressed in c.g.s. units.}
${\cal R} \sim 10^{-3} {\cal E}_{41}^{1/3} (E_{\rm LSD}/1~{\rm TeV})^{-4/3}$ cm.  Somewhat larger dimensions,
up to $\sim 0.1$-1 cm, are possible if the object is a GUT-scale `spring' loop.

\begin{figure}
\epsscale{1}
\plotone{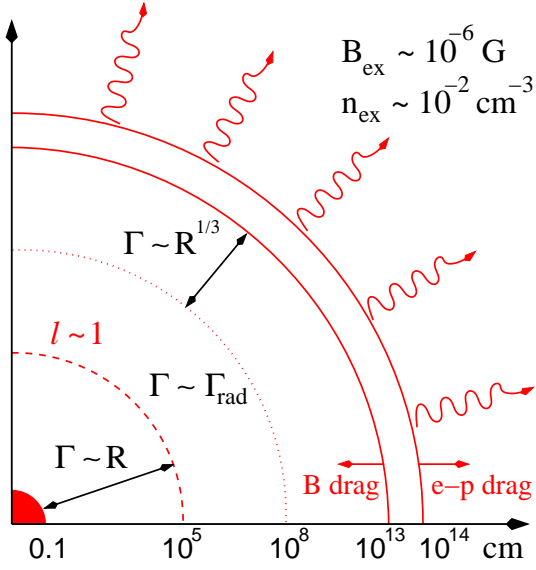}
\vskip .1in
\caption{Characteristic radii of acceleration and low-frequency electromagnetic emission
following a tiny electromagnetic explosion of size $\sim 1$ mm in an ISM-like medium.  
The explosion forms a thin shell with Lorentz factor increasing linearly with radius out
to the radius (\ref{eq:gamrad}) where the comoving radiation compactness drops below unity.
The internal Lorentz force induces further spreading outside the radius (\ref{eq:rl}).
Reflection of an ambient magnetic field into a superluminal electromagnetic wave damps
the motion of the shell at $R_s \sim 10^{13}$ cm (Equation (\ref{eq:rdecB2})), whereas 
electron drag becomes important at $\sim 10^{14}$ cm (Equation (\ref{eq:rdece2})),
for an ambient free electron density $n_{\rm ex} \sim 10^{-2}$ cm$^{-3}$ and magnetic field
$B_{\rm ex} \sim 10^{-6}$ G.  Electron drag sets in first when $n_{\rm ex} \gtrsim 0.1$ 
cm$^{-3}$, that is, when the ambient Alfv\'en speed is smaller than Equation 
(\ref{eq:alfcrit}).}
\vskip .2in
\label{fig:accel}
\end{figure}

In this section we treat the acceleration, deceleration, and radial spreading of this expanding shell.
Figure \ref{fig:accel} summarizes the processes involved.
Initially all embedded particles experience a large optical depth to scattering and absorption.  The
final transition to photon transparency, caused by the annihilation of electron-positron pairs, has a 
qualitative impact on the shell dynamics.  But the initial acceleration phase is insensitive to the relative
proportions of thermally excited particles and long-range electromagnetic fields.

These relative proportions are uncertain.  Only a small density of charged particles 
is needed to support electromagnetic field gradients on a scale ${\cal R}$.
This allows a cold beginning to the explosion:  the `charge-starvation' density is suppressed compared with 
$(E_{\rm LSD}/\hbar c)^3$ by a factor $(E_{\rm LSD} {\cal R} /\hbar c)^{-1} \sim  
10^{-16} (E_{\rm LSD}/{\rm TeV})^{-1} {\cal R}_{-1}^{-1}$.  In addition, it has been argued that
the QCD vacuum may become superconducting due to the condensation of charged meson in the presence of a sufficiently
strong magnetic field ($B \sim 0.6(e\hbar c)^{-1} ({\rm GeV})^2 = 10^{20}$ G: \citealt{chernodub2010}). 
On the other hand, a cascade process could transfer significant energy from long-range fields to thermal plasma,
as appears to happen in magnetar flares \citep{tb98}.

Hence, in what follows, we simply normalize the thermal energy as
\be
{\cal E}_{\rm th} = \varepsilon_{\rm th} {\cal E},
\ee
considering both $\varepsilon_{\rm th} \sim 0.1-1$ (a hot magnetized shell)
and $\varepsilon_{\rm th} \lesssim 10^{-6}$ (a cool shell).  Radio photons
are emitted by different channels in these two regimes.

\subsection{Prompt Phase}

The initial expansion profile is simple, with 
shell Lorentz factor $\Gamma_s \sim R_s/{\cal R}$ (Figure \ref{fig:accel}).
The thermal particles collimate radially as they expand, pulling
the magnetic field with them \citep{rt13a,rt13b,gt14}.  Both
hot and cool magnetized fireballs experience this rapid, prompt
acceleration, as long as the thermal (photon and relativistic pair) energy
dominates the rest energy.
Magnetized outflows in which non-relativistic material dominates the matter stress 
accelerate more slowly, $\Gamma_s \propto R_s^{1/3}$ \citep{drenkhahn02,granot11}.

The Poynting flux vanishes in a frame moving outward radially with Lorentz factor $\Gamma_s$.
In this frame, the embedded photons have a nearly isotropic distribution, and the long-range
electromagnetic field is purely magnetic, 
\be
B' = \left({{\cal E_P}\over R_s^2\cdot \Gamma_s {\cal R}}\right)^{1/2} \sim 
\left({{\cal E_P}\over R_s^3}\right)^{1/2}.
\ee
Breakdown of the magnetohydrodynamic (MHD) approximation only occurs on a very large scale, if at all (Section \ref{s:mhd}).

The simple prompt expansion law allows us to focus on an intermediate stage
where photons and $e^\pm$ pairs are the only thermally created particles remaining.
We ignore changes in the number of relativistic
degrees of freedom during the expansion, and the loss of energy associated with transparency to neutrinos.
The fireball is also assumed to have vanishing net baryon and lepton number.

The energy divides into the respective components as
\ba\label{eq:etot}
{\cal E} &=& {\cal E}_P + {\cal E}_\gamma + {\cal E}_\pm \nn 
&\simeq& 4\pi R_s^2\Delta R_s \left[{B^2\over 4\pi} + {4\Gamma_s^2\over 3}U_\gamma' + \Gamma_s^2U_\pm'\right].
\ea
Here $R_s$ is the shell radius, $U_\gamma'$ is the comoving photon energy density
and $U_\pm' \simeq n_\pm' m_ec^2$ during freeze-out of pairs.  
We combine the integral (\ref{eq:etot}) with conservation of magnetic flux and entropy,
\be\label{eq:flux}
\Gamma_s B' R_s \Delta R_s = {\rm const}; \quad \Gamma_s (U_\gamma')^{3/4} R_s^2 \Delta R_s = {\rm const}.
\ee
One finds that $\Gamma_s(R_s)$ increasing linearly in $R_s$ is the only power-law solution when i) $U_\gamma' \gg U_\pm'$ and ii) ${\cal R} \gg R_s/\Gamma_s^2$ (so that the shell is not able
to spread radially).

There is a useful comparison here with a cold magnetized shell.
Equations (\ref{eq:etot}) and (\ref{eq:flux}) imply in that case that the shell expands with 
constant $\Gamma_s$ in spherical geometry.  
Internal pressure gradients within the shell allow $\Gamma_s(t) \propto t^{1/3}$ in {\it planar} geometry,
which therefore is the dominant scaling in spherical geometry as well, $\Gamma_s(R_s)
\propto R_s^{1/3}$ \citep{granot11,lyutikov10}.  Note that the latter effect does not depend on dissipation.  As long as the advected rest energy dominates the thermal energy, dissipation must be introduced to accelerate a steady magnetized outflow
\citep{drenkhahn02}.

In summary, even a steady and spherical but hot
magnetized wind will accelerate rapidly, $\Gamma_s(R_s) \propto R_s$,
independent of the ratio of Poynting flux to relativistic thermal energy flux.  
This means that internal spreading of our thin shell has a negligible effect during the
prompt acceleration phase; but still gives a significant {\it late} boost
to the shell after photon pressure weakens (Section \ref{s:internal}).

\subsection{Coupling between Photons and Magnetized Pairs}

A simple physical picture emerges when cold pairs are included in the dynamical evolution of the shell,
as they must be when the comoving temperature drops below $\sim m_ec^2/\kB$.  
Even though ${\cal E}_P \gg {\cal E}_\pm$, the pairs dominate the radial inertia of the shell
because $\Gamma_s \gg (\sigma_\pm')^{1/3}$.  Here
\be
\sigma_\pm' \equiv {(B')^2\over 4\pi n_\pm' m_ec^2}
\ee 
is the comoving magnetization.  In this situation, the radial flow is faster than
a fast magnetosonic wave, and changes in the momentum flux carried by the
electromagnetic field are suppressed by a factor $\Gamma_s^{-3}$ (e.g. \citealt{gj70}).  

The prompt linear phase of acceleration, $\Gamma_s \sim R_s/{\cal R}$, ends when the anisotropic pressure of the photons
drops below a critical level.   This transition occurs when the comoving `compactness',
\be
\ell' = {\sigma_T U_\gamma' (R_s/\Gamma_s)\over m_ec^2} \sim {3\sigma_T {\cal E}_\gamma {\cal R}^2\over 16\pi r^4 m_ec^2},
\ee
drops below unity.   The shell Lorentz factor (temporarily) saturates at a value $\Gamma_{\rm rad}$ at a radius
$R_{\rm rad}$ given by
\ba\label{eq:gamrad}
\Gamma_{\rm rad} &=& 8.3\times 10^5 {\cal E}_{\gamma,41}^{1/4} {\cal R}_{-1}^{-1/2}; \nn
R_{\rm rad} &=& \Gamma_{\rm rad} {\cal R} = 8.3\times 10^4 {\cal E}_{\gamma,41}^{1/4} {\cal R}_{-1}^{1/2}\;{\rm cm}.
\ea

At this point the comoving temperature has dropped below $m_ec^2$ and the annihilation of pairs has
frozen out.  The kinetic energy of the residual pairs comprises a much larger fraction of the outflow
energy than it does in a gamma-ray burst outflow \citep{paczynski90,shemi90}, and may even
reach a significant fraction of the total before deceleration begins.  This is due to the relatively low compactness 
\be\label{eq:ell}
\ell' = 7.6\times 10^5\left({\kB T'\over m_ec^2}\right)^4 {\cal R}_{-1}
\ee
when the comoving temperature reaches $\kB T' \sim m_ec^2$, as compared with a fireball source of dimension 
${\cal R} \gtrsim 10^6$ cm.  

The pairs freeze out when $\tau_T' \equiv \sigma_T n_\pm' (R_s/\Gamma_s)  \sim \sigma_T c/\langle v\sigma_{\rm ann}\rangle
= 16/3$, that is, when optical depth to annihilation passes through unity.  Pairs with vanishing chemical potential
supply $\tau_T' \sim 0.2 \ell'$$(\kB T'/m_ec^2)^{-5/2}e^{-m_ec^2/\kB T'}$, which together with Equation (\ref{eq:ell})
gives $\kB T_f' \simeq m_ec^2/7$.  The net mass in pairs advected outward by the shell is
\be
{M_\pm c^2\over {\cal E}_\gamma} = {4\pi R_s^2 \Delta R_s \Gamma_s n_\pm' m_e\over {\cal E}_\gamma} 
 \sim {4\over \Gamma_{s,f}\ell'} \sim {0.02\over \Gamma_{s,f} {\cal R}_{-1}}.
\ee
Here $\Gamma_{s,f}$ is smaller than $\Gamma_{\rm rad}$ (Equation (\ref{eq:gamrad})) by a factor $\sim 0.25$,
corresponding to 
\be
{M_\pm c^2\over {\cal E}_\gamma} = 7\times 10^{-8} {\cal E}_{\gamma,41}^{-1/4} {\cal R}_{-1}^{-1/4}.
\ee

At this point in the expansion the advected magnetic field is very nearly non-radial.  The magnetization has decreased to 
\be
\sigma'_\pm (R_{\rm rad}) \sim {\Gamma_{\rm max}\over \Gamma_{\rm rad}},
\ee
where
\be\label{eq:gamax}
\Gamma_{\rm max} = {{\cal E}\over M_\pm c^2} = 7.8\times 10^7 \varepsilon_{{\rm th},-1}^{-3/4} 
      {\cal E}_{41}^{1/4} {\cal R}_{-1}^{1/4}
\ee
is the maximum achievable Lorentz factor obtained by balancing the long-range electromagnetic energy
(as measured during an early stage of the expansion) with the kinetic energy of
the frozen pairs.  Hereafter we normalize ${\cal E}_\gamma$ to a fixed fraction $\varepsilon_{\rm th}$ of the total energy.

\subsection{Delayed Acceleration by the Internal Lorentz Force}\label{s:internal}

Slower acceleration continues beyond the terminal radius $R_{\rm rad}$ of radiatively driven acceleration,
and is sourced by the radial spreading of the magnetic field.   This process begins even before the shell regains causal
contact in the radial direction (e.g. Granot et al. 2011),
\ba\label{eq:gammhd}
\Gamma_s(r) &\sim& \Gamma_{\rm rad}\left[\sigma'_\pm(R_{\rm rad}) {r\over 2\Gamma_{\rm rad}^2 {\cal R}}\right]^{1/3}\nn
&=& \Gamma_{\rm  max}\left({r\over 2\Gamma_{\rm max}^2 {\cal R}}\right)^{1/3};\quad R_{\rm L} < r < R_{\rm sat}.\nn
\ea
This process begins around the radius
\be\label{eq:rl}
R_{\rm L} = 2(\Gamma_{\rm rad}^3/\Gamma_{\rm max}){\cal R} = 2.6\times 10^8 \varepsilon_{\rm th,-1}^{3/2} 
    {\cal E}_{41}^{1/2} {\cal R}_{-1}^{-3/4}\quad {\rm cm}.
\ee
The magnetization drops to $\sigma'_\pm \sim 1$ and the kinetic energy of the pairs comprises $\sim {1\over 2}$
of the total shell energy at a radius
\be\label{eq:rsat}
R_{\rm sat} = 2\Gamma_{\rm max}^2 {\cal R} \sim 1.2\times 10^{15} \varepsilon_{{\rm th},-1}^{-3/2} {\cal E}_{41}^{1/2} {\cal R}_{-1}^{3/2}
\;{\rm cm}.
\ee
The shell dynamics inside this radius can be modified by its interaction with the external medium, which is
examined in the following section.

\section{Interaction with a Magnetized and \\ Ionized Medium}\label{s:decel}

The shell experiences drag by i) reflecting an ambient magnetic field into a
propagating electromagnetic wave (see also \citealt{blandford77}); ii) 
deflecting ambient electrons \citep{noerd71}; and iii) through 
an electrostatic coupling between radially polarized electrons and ions.   The relative 
importance of these forces depends on ${\cal E}$ and $\varepsilon_{\rm th}$, as well as on ambient conditions.
The shell energy is converted with a relatively high efficiency to a low-frequency electromagnetic pulse if
the drag force i) dominates.   Direct conversion of the advected magnetic field to a propagating superluminal
pulse may also occur, before the shell decelerates significantly, if the ambient medium is very rarefied
(Section \ref{s:em}).

\subsection{Deflection of Ambient Electrons and Ions}\label{s:deflect}

A particle of charge $q$ traversing a non-radial magnetic flux\footnote{To simplify the notation, we choose
a coordinate system in which $\Phi_B$ and $A_\perp$ are both positive within the shell, in a gauge
where $A_\perp = 0$ outside the shell.}
 $\Phi_B(r) = \int^{R_s}_r B dr$ (between the forward edge of the shell and radius $r < R_s$)  absorbs 
transverse momentum $p_\perp = -q\Phi_B(r)/c$.  This relation can be obtained either by integrating the
Lorentz force; or by invoking conservation of the generalized transverse momentum $p_\perp + qA_\perp/c$,
where the vector potential $A_\perp = \Phi_B$.   Hereafter we choose a negative charge and use $e$ to denote 
the magnitude of the electron charge.

The strength of the interaction between electron and
shell is characterized by the dimensionless parameter
\be
{|p_\perp|\over m_e c} = {e\Phi_B\over m_ec^2} \equiv a_{e\perp}.
\ee
Integrating radially through a uniformly magnetized shell, there is a net shift in the vector potential,
\be\label{eq:dela}
\Delta a_{e\perp} = {e({\cal E} {\cal R})^{1/2}\over m_ec^2 r}.
\ee
At a finite depth $\xi = r-R_s < 0$ behind the front of the shell,
\be
a_{e\perp}(\xi) = \Delta a_{\perp} {|\xi|\over \Delta R_s}.
\ee
In the absence of external charges, the electrons overlapping the shell gain relativistic energies
($\Delta a_{e\perp} > 1$) out to a radius
\be
R_{\rm rel} \sim 5.9\times 10^{16}\,{\cal E}_{41}^{1/2} {\cal R}_{-1}^{1/2}\quad {\rm cm}.
\ee
As we show in Section \ref{s:hf}, this transition is pushed to a significantly smaller radius
for a superluminal electromagnetic wave when the effects of plasma dispersion are taken into account.

Electrons and ions flowing through the shell do not absorb
net transverse momentum as long as the ambient medium is charge neutral.  However, a longitudinal polarization is
established within the shell due to the much greater angular deflection of the electrons.  (This effect was not
taken into account by \cite{noerd71}.)  In the frame of the shell, an electron gains a Lorentz factor
\be\label{eq:gamrest}
\gamma'_e = \Gamma_s\left[1-{e\Phi\over m_ec^2}\right] \equiv \Gamma_s(1+\phi).
\ee
Here $\Phi < 0$ is the electrostatic potential as measured in the `ambient' frame (the rest frame of the ambient medium).
The strength of this radial polarization increases with the shell thickness for given ${\cal E}$.   A proton gains
an energy $\gamma_p' = \Gamma_s[1-(m_e/m_p)\phi]$.  

The energy imparted to a swept-up electron is obtained by a Lorentz transformation back to the ambient frame,
\ba\label{eq:game}
\gamma_e &=& \Gamma_s(\gamma'_e - \beta_s\beta'_e\gamma'_e) \nn
       &\simeq& {a_{e\perp}^2\over 2(1+\phi)} \quad\quad (\Gamma_s \gg a_{e\perp} \gg 1).
\ea
Here $\beta$ is the radial speed in units of $c$.   

A forward section of the shell threaded by flux $\Phi_B$ will lose
a net radial momentum $\simeq \gamma_e(\Phi_B)m_ec$ to each transiting electron.  
Traversing a medium of free electron density $n_{\rm ex}$ through a distance $dr$, the change
in radial momentum is given by 
\be
\delta N_\pm m_ec \cdot d(\Gamma_s\beta_s) = - {e^2n_{\rm ex}\over 2m_ec^2} {\Phi_B^2\over 1+\phi} dr.
\ee
Here $\delta N_\pm = n_\pm (R_s -r) = n_\pm |\xi|$ is the column of embedded $e^\pm$, which dominate
the radial inertia.
Assuming that the shell is in radial dynamical equilibrium, the net force per unit solid angle is
\be\label{eq:drageq}
{d\over dt}\left({dM_\pm\over d\Omega} \Gamma_s\beta_s c\right) \;=\;
      - \gamma_e(\Delta a_{e\perp}) m_ec \cdot R_s^2 n_{\rm ex} c.
\ee
The front end of the shell feels much weaker drag if the non-radial magnetic field maintains a 
consistent sign throughout the shell.

The radial deflection of ambient electrons penetrating the shell can be written down similarly.
Integration of the radial momentum equation gives (Appendix \ref{a:strongem})
\be\label{eq:betae}
\gamma_e(1-\beta_e) = 1 + \phi 
\ee
This is easily checked in the regime $\Gamma_s \gg a_s \gg 1$ by boosting
from the shell rest frame, as above.  The penetrating electrons are pushed outward to a radial speed
\be
1-\beta_e \simeq {2(1+\phi)^2\over a_{e\perp}^2}\quad\quad (\Gamma_s \gg \gamma_{\parallel,e} \gg 1),
\ee
corresponding to a radial boost $\gamma_{\parallel,e} = (1-\beta_e^2)^{-1/2} \simeq a_{e\perp}/2(1+\phi)$.

We notice also that the electron energy is independent of the shell Lorentz factor as long as $\Gamma_s \gg a_{e\perp}$.  
This means that the same expressions hold when ambient charges interact with a 
large-amplitude superluminal wave -- even though in this case there is no frame in which the transverse electric field
vanishes.

The proton energy and drift speed are obtained from Equations (\ref{eq:game}) and (\ref{eq:betae})
by setting $\{a_{e\perp},\phi\} \rightarrow -(m_e/m_p)\{a_{e\perp},\phi\}$.  This means that $\gamma_p$ has a singularity
as $\phi \rightarrow m_p/m_e$.  This corresponds to a strong electrostatic exchange of radial momentum between
electrons and ions:  if $\phi$ reaches this threshold,  then the positive and negative charges begin to move
collectively with the same mean radial speed.  Further details, along with expressions valid also for 
$a_{e\perp} \ll 1$ and $a_{e\perp} < \phi$,  can be found in Appendix \ref{a:strongem}.

\subsubsection{Radial Electrostatic Field}\label{s:potential}
 
We now consider the feedback of radial polarization on the energization of charges 
transiting a thin, relativistic, magnetized shell.  For simplicity, the ambient 
plasma is assumed to be composed entirely of H$^+$ (protons) and the compensating free electrons.  

The electrostatic potential is obtained from Gauss' law, which for a thin shell reads
\be\label{eq:gauss}
-{\partial^2\phi\over\partial r^2} = 4\pi e (n_p - n_e) = 
 4\pi e n_{\rm ex}\left({1\over 1-\beta_p} - {1\over 1-\beta_e}\right).
\ee
(The second equality follows from the assumption of a quasi-steady particle flow across the shell.)
The right-hand side of Equation (\ref{eq:gauss}) can be expressed in terms of $\phi$ using Equation
(\ref{eq:betae}).
Then multiplying by $-\partial\phi/\partial r$ and integrating by parts gives
\be\label{eq:phieq}
\left({\partial\phi\over\partial r}\right)^2 = 
\left({a_{e\perp} \omega_{P,\rm ex}\over c}\right)^2\left[{\phi\over 1+\phi} + \left({m_e\over m_p}\right)^2
{\phi\over 1 - (m_e/m_p)\phi}\right],
\ee
where $\omega_{P,\rm ex} = (4\pi n_{\rm ex}e^2/m_e)^{1/2}$ is the ambient cold plasma frequency.   
Near the outer boundary of the shell, where $\phi \ll m_p/m_e$, this simplifies to
\be\label{eq:philin}
\phi(r) \simeq a_{e\perp}(r) {\omega_{P,\rm ex}(R_s-r)\over 2c}.
\ee
Deeper in the shell (if it is thick enough) the electrons and ions equilibrate at a mean speed
\be\label{eq:beta}
1 - \beta_e = 1-\beta_p = {2\over 3} \left({m_p\over a_{e\perp} m_e}\right)^2 \quad\quad (a_{e\perp} \gg m_p/m_e),
\ee
corresponding to a bulk Lorentz factor 
\be\label{eq:gampar}
\gamma_\parallel = {\sqrt{3}m_e\over 2m_p} a_{e\perp}.
\ee
Then the potential averages to $\langle\phi\rangle \simeq m_p/2m_e$, and the particle energies to
\be\label{eq:phiav}
\langle \gamma_e\rangle \simeq {3m_e\over 4m_p}a_{e\perp}^2 =  {m_p\over m_e}\langle\gamma_p\rangle.
\ee

The effects of radial polarization are important mainly when the ambient medium is very dense compared with the local ISM,
or when the shell is much broader than a radio wavelength, e.g. ${\cal R} \gg 10$ cm.  In the linear 
approximation (\ref{eq:philin}), the net potential drop across a shell of thickness $\Delta R_s \geq {\cal R}$ is
\ba\label{eq:rphi}
{\Delta\phi(R_s)\over m_p/m_e} &=& {\Delta R_s\over {\cal R}}{R_\phi\over R_s};\nn
        R_\phi &=& 3.0\times 10^6 \; (n_{\rm ex,0}\,{\cal E}_{41}\,{\cal R}_{-1}^3)^{1/2}\quad {\rm cm}.
\ea

\subsubsection{Transit of the Shell by Ambient Charges}
                                                              
Ambient electrons entirely penetrate the expanding shell only after it has expanded well beyond the radius
$R_{\rm rad}$ where radiation-driven acceleration is complete.
Setting $\phi \rightarrow 0$, corresponding to ISM-like electron density, the net radial displacement 
$\xi = r - R_s$ increases as
\be
\left|{d\xi\over dr}\right| \simeq 1-\beta_e \simeq {2\over a_{e\perp}^2}. 
\ee
Taking $a_{e\perp} = \Delta a_{e\perp} \cdot \xi/{\cal R}$, this integrates to $\xi = {\cal R}$ at a radius
\be
R_{{\rm trans},e} = \left[{e^2{\cal E}\, {\cal R}^2\over 2(m_ec^2)^2}\right]^{1/3} = 5.6\times 10^{10}\,{\cal E}_{41}^{1/3}
{\cal R}_{-1}^{2/3}\quad{\rm cm}.
\ee
The corresponding transit radius for ions of mass $m_A = Am_pc^2$ and charge $Ze$ is
\ba\label{eq:rtransi}
R_{{\rm trans},i} &=& \left({Z\over A}{m_e\over m_p}\right)^{2/3}\,R_{{\rm trans},e}\nn
                &=& 3.7\times 10^8\,\left({Z\over A}\right)^{2/3}{\cal E}_{41}^{1/3}{\cal R}_{-1}^{2/3}\quad{\rm cm}.
\ea
Comparing with Equation (\ref{eq:rphi}), one sees that the radial polarization
must be taken into account when $R_{{\rm trans},e} \lesssim (m_p/m_e)R_\phi$,
corresponding to an ambient density $n_{\rm ex} \gtrsim 10^2\,{\cal E}_{41}^{-1/3}{\cal R}_{-1}^{-5/3}$ cm$^{-3}$.

\subsection{Interaction with an External Magnetic Field}\label{s:Breflect}

A relativistic conducting shell moving through a magnetized medium also feels a drag force 
from the reflection of the ambient magnetic field ${\bf B}_{\rm ex}$ into a propagating electromagnetic wave.  
We focus here on the regime where the comoving plasma frequency $\omega_{P\pm}'$ of the advected
pair gas satisfies $\omega_{P\pm}' \gg c/\Delta R_s' = c/\Gamma_s\Delta R_s$.  Then the shell behaves as
a nearly perfect conductor in response to a low-frequency electromagnetic perturbation.  

The electromagnetic field outside the shell has ingoing ($-$) and outgoing ($+$) components, the first
representing the Lorentz-boosted external magnetic field,
\ba
{\bf B}'_\perp &\;=\;& {\bf B}'_{\perp-} + {\bf B}'_{\perp+} \;=\;
      \Gamma_s({\bf B}_{\rm ex} - \hat{r}\cdot{\bf B}_{\rm ex})  + \hat{r}\times{\bf E}_{\perp+}'\nn
{\bf E}'_\perp &\;=\;& {\bf E}'_{\perp-} + {\bf E}'_{\perp+} \;=\;
        \Gamma_s\beta_s \hat{r}\times{\bf B}_{\rm ex} + {\bf E}'_{\perp+}.
\ea
The comoving transverse electric field vanishes at the shell surface, giving ${\bf E}'_{\perp-} \simeq
     -\Gamma_s \hat{r}\times{\bf B}_{\rm ex}$.  The radial force acting per unit solid angle of shell is
\be
{d\over dt'}\left({dM_\pm\over d\Omega} \beta_s' c\right) \;=\;
- R_s^2\left({{\bf E}'_{\perp-}\times{\bf B}'_{\perp-}\over 4\pi} + 
        {{\bf E}'_{\perp+}\times{\bf B}'_{\perp+}\over 4\pi}\right).
\ee
The radial force is invariant under a radial Lorentz boost.  Substituting for the field components then
gives
\ba\label{eq:magdrag}
{d\over dt}\left({dM_\pm\over d\Omega} \Gamma_s\beta_s c\right) \;&=&\;
- 2R_s^2\Gamma_s^2{B_{\rm ex}^2-(\hat{r}\cdot{\bf B}_{\rm ex})^2\over 4\pi} \nn
&\equiv& - 2R_s^2\Gamma_s^2{B_{\rm ex\perp}^2\over 4\pi}.
\ea
Here $B_{\rm ex\perp}$ is the component of the external magnetic field perpendicular to the shell velocity.
An equivalent global expression describing the interaction of a spherical shell with a uniform magnetic field
is derived by \cite{blandford77}.

\subsection{Characteristic Radii}

The dominant drag force varies with ambient conditions and shell properties.  Here we write down 
characteristic radii for the shell to decelerate.  There are two regimes of interest:
i) the deceleration is delayed to a radius exceeding $R_{\rm sat}$ (Equation (\ref{eq:rsat})), where
most of the shell energy is converted to kinetic energy of the embedded pairs; and ii) the shell decelerates
while the internal Lorentz force is still acting on it, so that
$\Gamma_{s0}$ is smaller than the kinematic limit $\Gamma_{\rm max}$ (Equation (\ref{eq:gamax})).

\subsubsection{Drag -- ISM Conditions}\label{s:lowrho}

The simplest case is slow deceleration into an ambient medium characteristic of the ISM
of galaxies:  $n_{\rm ex} \sim 0.01$-$1$ cm$^{-3}$ and $B_{\rm ex} \sim 1$-10 $\mu$G.  
Then the radial polarization of the shell can be neglected. 

When electron deflection dominates the drag, we have
\be
\Gamma_s = \Gamma_{s0} - {2\pi n_{\rm ex} e^2 {\cal R}\over m_ec^2}\Gamma_{\rm max} R_s.
\ee
Here we treat the shell as a single dynamical unit; $\Gamma_{s0}$ is the shell Lorentz factor in the 
absence of drag.  Then $\Gamma_s$ drops to one half the drag-free value at the radius
\ba\label{eq:rdece}
R_{{\rm dec},e}(\Gamma_{s0}) &=& {\Gamma_{s0}\over \Gamma_{\rm max}} {m_ec^2\over 4\pi n_{\rm ex} e^2 {\cal R}} \nn
         &=& 2.8\times 10^{12}\;{\Gamma_{s0}\over\Gamma_{\rm max}} {\cal R}_{-1}^{-1}n_{\rm ex,0}^{-1} \quad{\rm cm}.\nn
\ea
The slow action of the internal Lorentz force allows us to substitute Equation (\ref{eq:gammhd}) for $\Gamma_{s0}$,
and then re-solve for the deceleration radius:
\be\label{eq:rdece2}
R_{{\rm dec},e} = 1.4\times 10^{11}\, n_{\rm ex,0}^{-3/2}\varepsilon_{\rm th,-1}^{3/4}{\cal E}_{41}^{-1/4} 
                   {\cal R}_{-1}^{-5/2}\quad {\rm cm}.
\ee

Drag by the external magnetic field is imposed at the outer boundary of the shell, rather than
volumetrically as in the case of electron drag.  If the magnetic drag mainly acts outside the saturation
radius (\ref{eq:rsat}), where the kinetic energy of the embedded pairs rises to a significant fraction of
the explosion energy, then
\be\label{eq:gammag}
{1\over \Gamma_s} - {1\over\Gamma_{s0}} = {2R_s^3 B_{\rm ex\perp}^2\over 3{\cal E}}\Gamma_{\rm max}.
\ee
The deceleration radius is 
\be\label{eq:rdecB}
R_{{\rm dec},B} = \left({3{\cal E}\over 4\Gamma_{\rm max}^2 B_{\rm ex\perp}^2}\right)^{1/3}.
\ee
This expression only applies if the external magnetic field is relatively weak and the thermal
content of the explosion is high.  That is because the condition $R_{{\rm dec},B} > 2\Gamma_{\rm max}^2{\cal R}$ 
corresponds to
\be\label{eq:Bmax}
B_{\rm ex\perp} < 1\times 10^{-8} \left({\varepsilon_{\rm th}\over 0.5}\right)^3 {\cal E}_{41}^{-1/2}
                       {\cal R}_{-1}^{-5/2} \quad {\rm G}.
\ee
A treatment of magnetic drag in more realistic ISM conditions is given in Section \ref{s:diffaccel}.

\begin{figure}
\epsscale{1}
\plotone{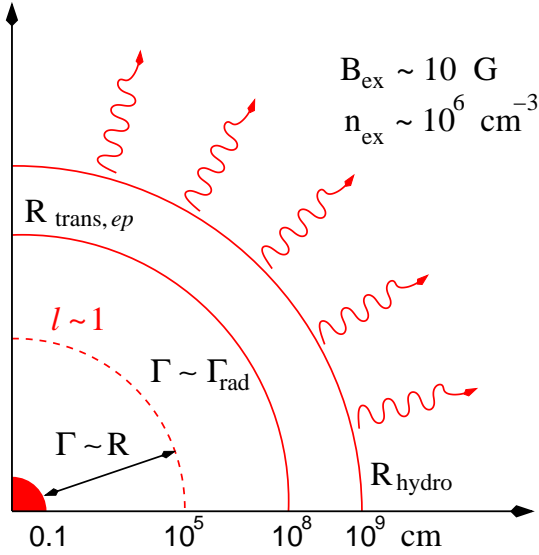}
\vskip .1in
\caption{Interaction of a tiny electromagnetic explosion with a high-density medium, such
as is encountered in a RIAF near a SMBH.  Ions swept up by the expanding magnetized shell
continue to surf the electromagnetic field out to the radius (\ref{eq:rtransep}).
The energy imparted to the ions during the first part of the expansion is recaptured by the
shell after it reaches the radius (\ref{eq:rdecp2}), whereupon it transitions to a hydromagnetic
explosion.  The interaction with an ambient magnetic field containing a modest fraction of the 
ambient rest energy is an efficient driver of a low-frequency electromagnetic wave.}
\vskip .2in
\label{fig:accel2}
\end{figure}

\subsubsection{Drag -- High Density Medium}\label{s:highrho}

LSDs may be trapped in long-lived orbits around supermassive black holes (Paper II).  Here the electron
density (magnetic energy density) may easily reach $\sim 10^8$ times ($\sim 10^{14}$ times) the local
ISM value -- even when the black hole is a relatively slow accretor like the Galactic Center
black hole \citep{yuan03}.  

Now the interaction with ambient free electrons and ions is the dominant source of drag.  Ions
swept up by the expanding electromagnetic shell eventually re-energize it and convert it to a hydromagnetic
explosion (Figure \ref{fig:accel2}).

The radial polarization of the relativistic shell must be taken into account when evaluating the
drag force from penetrating charges.   We first consider the case where  $\phi$ is smaller than the 
limiting value $m_p/m_e$.    Then Equation (\ref{eq:phieq}) integrates to give 
\be
\phi(\xi) \simeq \Delta a_{e\perp}{\omega_{P,\rm ex}\xi^2\over 2\Delta R_s c}.
\ee
Substituting this into Equation (\ref{eq:game}) gives
\be
\gamma_e = \Delta a_{e\perp} {c\over \omega_{P,\rm ex}\Delta R_s},
\ee
which is independent of depth $\xi$ in the shell.  
Further substituting for $\gamma_e$ in Equation (\ref{eq:drageq}) gives
\be
\Gamma_s = \Gamma_{s0} - {\omega_{P,\rm ex} m_e c\over 2e({\cal E}{\cal R})^{1/2}}\Gamma_{\rm max} R_s^2.
\ee
This expression holds inside the polarization radius
\ba
R_{s,\rm pol,1} &=& \left[\pi e^4 n_{\rm ex} {\cal E} {\cal R}^3 \over (m_ec^2)^2\right]^{1/2} \nn
                &=& 5.5\times 10^{11}\; {\cal E}_{41}^{1/2} {\cal R}_{-1}^{3/2} n_{\rm ex,4}^{1/2}\quad {\rm cm}.
\ea
The deceleration radius is obtained by substituting Equation (\ref{eq:gammhd}) for $\Gamma_s$, giving
\be\label{eq:rdece3}
R_{{\rm dec},e} = 1.9\times 10^9\;\varepsilon_{\rm th,-1}^{3/10} {\cal E}_{41}^{1/5} n_{\rm ex,4}^{-3/10} \quad {\rm cm}.
\ee

An even denser ambient medium allows $\phi$ to reach $\sim m_p/m_e$.
Now ions and electrons attain energy equipartition within the shell, and Equation (\ref{eq:drageq})
is replaced with
\be\label{eq:drageq2}
{d\over dt}\left({dM_\pm\over d\Omega} \Gamma_s\beta_s c\right) \;=\;
      - 2\langle\gamma_e\rangle m_ec \cdot R_s^2 n_{\rm ex} c,
\ee
where $\langle\gamma_e\rangle$ is given by Equation (\ref{eq:phiav}).  We now have
\be\label{eq:gamspart}
\Gamma_s = \Gamma_{s0} - {6\pi n_{\rm ex}e^2 {\cal R}\over m_pc^2} \Gamma_{\rm max} R_s,
\ee
with
\be\label{eq:rdecep}
R_{{\rm dec},ep} = 2.0\times 10^9\; \varepsilon_{\rm th,-1}^{3/4} {\cal E}_{41}^{-1/4} {\cal R}_{-1}^{-9/4}
n_{\rm ex,4}^{-3/2} \quad {\rm cm}.
\ee
These expressions only apply inside the radius
\be
R_{s,\rm pol,2} \simeq {3m_e\over m_p}R_{s,\rm pol,1}
                = 9.0\times 10^8\; {\cal E}_{41}^{1/2} {\cal R}_{-1}^{3/2} n_{\rm ex,4}^{1/2}\quad {\rm cm}.
\ee
Note that the particle drag radius is given by the minimum of
Equation (\ref{eq:rdece3}), corresponding to $1 < \phi < m_p/m_e$, and Equation (\ref{eq:rdecep}),
corresponding to strong electrostatic coupling of the swept-up electrons and ions.  The transition between
these two drag regimes occurs at $n_{\rm ex} \sim 10^4$ cm$^{-3}$.

The deceleration radius $R_{{\rm dec},ep}$ as given by Equation (\ref{eq:rdecep}) must lie outside the
transit radius of the swept-up electrons and ions.  This transit radius is obtained by equating Equation 
(\ref{eq:gampar}) with the shell Lorentz factor ($\Gamma_{\rm rad}$), corresponding to
\be\label{eq:rtransep}
R_{{\rm trans},ep} = 6\times 10^7 \, \varepsilon_{\rm th,-1}^{-1/4}{\cal E}_{41}^{1/4} {\cal R}_{-1}\quad {\rm cm}.
\ee
Otherwise the electromagnetic shell enters a hydromagnetic regime which we now describe.

We now emphasize an important distinction between the interaction of the shell with the
ambient magnetic field, and its interaction with charged particles.
Whereas reflection of the magnetic field causes a permanent loss of energy
to a superluminal wave outside the shell, the ions which transit the shell collect 
{\it behind} it and eventually re-energize it.  (The electrons however do not have the same effect:
their kinetic energy, which may approach that of the ions in the highest density regime, 
is efficiently radiated away:  see Equation (\ref{eq:rrad}).)  The cutoff in Lorentz factor implied by Equation
(\ref{eq:gamspart}) is sharp enough that the re-collision of most of the heated ions with the shell is located
just outside the transit radius, which means that $\Gamma_s$ maintains a plateau at $\sim \Gamma_{\rm rad}$.

The final expansion regime closely approximates a relativistic adiabatic blast familiar from the theory of gamma-ray
bursts, in which the swept-up particles are strongly scattered, with 
$\gamma_\perp \sim \gamma_\parallel \sim \Gamma_s$ \citep{rm92},  
\be\label{eq:gamfin}
\Gamma_s(R) \simeq \left({3{\cal E}\over 4\pi \rho_{\rm ex}c^2R_s^3}\right)^{1/2}\quad (R> R_{\rm hydro}).
\ee
Here $\rho_{\rm ex} \simeq m_p n_{\rm ex}$ is the ambient plasma rest mass density.  The transition from 
$\Gamma_s \sim \Gamma_{\rm rad}$ to Equation (\ref{eq:gamfin}) occurs at the radius
\be\label{eq:rdecp2}
R_{\rm hydro} \sim 4\times 10^8\,\varepsilon_{\rm th,-1}^{-1/6}{\cal E}_{41}^{1/6}{\cal R}_{-1}^{1/3} 
                     n_{\rm ex,6}^{-1/3}\quad {\rm cm}.
\ee

\subsection{Equation of Motion of a \\ Thin, Relativistic, Magnetized Shell}\label{s:eom}

Here we collect the various forces acting on a thin, relativistic shell. 
Initially the shell is hot and pair-loaded, but its energy is dominated 
by a non-radial magnetic field.  The equation of
motion obtained here is used in Section \ref{s:radio} to calculate the long-term evolution of the shell, 
along with the emitted spectrum of low-frequency radiation.

The initial hot expansion phase is generally well separated from the deceleration phase
during which low-frequency emission takes place.  Nonetheless, we include for completeness
the radiation force acting on the embedded pairs.  Here it is sufficient to assume that
the pairs are cold and scattering is in the Thomson regime.  Then (e.g. \citealt{rt13a,rt13b})
\be
f_{\rm rad} = {\sigma_T\over 4\Gamma_s^2 c}\left[1-\left({\Gamma_s\over\Gamma_{\rm eq}(r)}\right)^4\right]
{{\cal E}_\gamma c\over 4\pi R_s^2 \Delta R_s}.
\ee
The last factor on the right-hand side is the radiation energy flux; the negative term inside
the brackets accounts for the effects of relativistic aberration.  The radiation force vanishes in a frame moving 
at Lorentz factor $\Gamma_{\rm eq} \simeq r/{\cal R}$, where ${\cal R}$ measures the volume of the energy release. 

Hydromagnetic acceleration due to radial spreading \citep{granot11} can be combined with 
other radial forces in the following
heuristic way.  The shell is divided into an inner component with very high (essentially infinite)
magnetization, and a second outer shell that carries the inertia of the pairs.   The inner shell has a
non-radial magnetic field as fixed by the conservation of non-radial magnetic flux,
\be
B_s = {\Delta A_\perp\over \Delta R_s} = { ({\cal E} {\cal R})^{1/2} \over R_s\Delta R_s },
\ee
where $\Delta A_\perp$ is the net displacement of the vector potential across the shell.
In the rest frame of the material shell (boosted by Lorentz factor $\Gamma_s$), the inner
shell transmits a Poynting flux
\be\label{eq:poynt}
{dS_{\rm P}'\over d\Omega} \;=\; R_s^2 {1-\beta_s\over 1+\beta_s} {B_s^2\over 4\pi}c \;\simeq\; 
                       {{\cal R} c\over 4(\Gamma_s\Delta R_s)^2}{d{\cal E}\over d\Omega} 
\ee
per unit solid angle.  The hydromagnetic acceleration of the shell is then computed from
\be\label{eq:mhdforce}
{d\over dt}\left(\Gamma_s\beta_s {dM_\pm\over d\Omega} c\right) = 
{d\over dt'}\left(\beta_s' {dM_\pm\over d\Omega} c\right) = {1\over c}{dS_{\rm P}'\over d\Omega}.
\ee
Combining Equations (\ref{eq:poynt}) and (\ref{eq:mhdforce}) reproduces the scaling (\ref{eq:gammhd}).

The effects of radiation pressure and internal MHD stresses can now be combined with the drag
forces described in Sections \ref{s:deflect} and \ref{s:Breflect} to give the equation of motion
\ba\label{eq:netforce}
{d\over dt}\left(\Gamma_s {dM_\pm\over d\Omega} c\right) &\;=\;&  {f_{\rm rad}\over m_e} {dM_\pm\over d\Omega}
\;+\; {{\cal R} \over 4(\Gamma_s\Delta R_s)^2}{d{\cal E}\over d\Omega} \nn
&\;-\;& R_s^2 n_{\rm ex} \gamma_e(\Delta a_{e\perp}) m_ec^2-2R_s^2\Gamma_s^2{B_{\rm ex\perp}^2\over 4\pi}.\nn
\ea
We have also
we have assumed that the entire shell is in radial causal contact, e.g. $R_s \gg \Delta R_s 
\cdot \Gamma_s^2$.   Deceleration may begin at an earlier stage, in which case 
the shell divides into distinct dynamical components (Section \ref{s:diffaccel}).

The radial force equation (\ref{eq:netforce}) can be written in a more transparent, 
dimensionless form using the characteristic radii (\ref{eq:rdece}) and (\ref{eq:rdecB}),
\ba\label{eq:netforce2}
{R_s\over\Gamma_{\rm max}}{d\Gamma_s\over dR_s} &=& {R_s\over 4\Gamma_s^2{\cal R}}
\left\{ \left({{\cal R}\over \Delta R_s}\right)^2 + {\ell_\gamma\over \Gamma_{\rm max}}
\left[1-\left({\Gamma_s\over\Gamma_{\rm eq}}\right)^4\right]\right\} \nn
&-& {3\over 2}\left({\Gamma_s\over\Gamma_{\rm max}}\right)^2\,\left[{R_s\over R_{{\rm dec},B}}\right]^3
- {R_s \over 2R_{{\rm dec},e}}.\nn
\ea
Here $R_{{\rm dec},e}$ is evaluated at $\Gamma_s = \Gamma_{\rm max}$, and
\be
\ell_\gamma \equiv {\sigma_T{\cal E}_\gamma\over 4\pi R_s^2 m_ec^2}
\ee
is the radiation compactness integrated radially through the shell. 

As $\Gamma_s$ drops below $\sim (R_s/{\cal R})^{1/2}$,  
we have to take into account the growth of $\Delta R_s$.  
Since $B'\Delta R_s' R_s$ and $n_\pm'\Delta R_s' R_s$ are both constants, 
the shell magnetization is
\be
\sigma_\pm' \simeq {\Gamma_{\rm max}\over\Gamma_s}\cdot{{\cal R}\over\Delta R_s}.
\ee
In the absence of external interaction, the magnetization reaches $\sigma'_\pm \sim 1$ at Lorentz 
factor $\sim \Gamma_{\rm max}$, and beyond that point the remaining electromagnetic energy is 
transferred to the embedded pairs.

Expansion of the shell in the comoving frame occurs at the fast mode speed
\be\label{eq:drshell}
{d\Delta R_s'\over dt'} = {c\over \Gamma_s} {d\Delta R_s'\over dr} \sim  c \left(1 + {1\over \sigma_\pm'}\right)^{1/2}.
\ee
In the case $\Gamma_s \simeq \Gamma_{\rm max} =$ const, the width grows to
\be\label{eq:delr}
{\Delta R_s\over {\cal R}} \sim \left({3R_s\over 2\Gamma_{\rm max}^2{\cal R}}\right)^{2/3};\quad
R_s > 2\Gamma_{\rm max}^2 {\cal R}.
\ee

\subsection{Slower Outer Shell}\label{s:diffaccel}

The simplifying assumption of a radially homogeneous shell must break down if the shell decelerates
at a radius $R_{\rm dec} \ll \Gamma_s^2 {\cal R}$.  Then a forward `contact layer' begins to decelerate before the
bulk of the shell.  Reconnection between the ambient and shell magnetic fields, which generally
are not aligned, can be expected to heat the particles near the contact.   This motivates a dynamical
model in which the motion of the contact is determined by a balance between the fluxes of momentum from
inside and outside.  The equation of motion (\ref{eq:netforce}) is then supplemented
with a coupled equation for the Lorentz factor $\Gamma_c$ of the contact layer.   

The contact layer has a rest mass $dM_{\pm,c}/d\Omega < dM_\pm/d\Omega$ and non-radial magnetic flux 
which grow as the faster shell material catches up from behind.  Neglecting the radiation force (which
is negligible at this stage)
\ba\label{eq:netforcec}
{d\over dt}\left(\Gamma_c {dM_{\pm,c}\over d\Omega} c\right) &\;=\;&  {{\cal R} \over 4(\Gamma_c\Delta R_s)^2}
{d{\cal E}\over d\Omega}  - 2R_s^2\Gamma_c^2{B_{\rm ex\perp}^2\over 4\pi}\nn
&\;-\;& R_s^2 n_{\rm ex} \gamma_e(a_{\perp,c}) m_ec^2.
\ea
Here 
\be
a_{\perp,c} = {dM_{\pm,c}/d\Omega\over dM_\pm/d\Omega}\,\Delta a_{e\perp}
\ee
measures the magnetic flux carried by the contact shell.  Its mass accumulates at a rate
\be\label{eq:dmcdt}
{1\over c}{d\over dt}{dM_{\pm,c}\over d\Omega} = {\beta_s-\beta_c\over\Delta R_s}{dM_\pm\over d\Omega}
\simeq {1\over 2\Delta R_s}\left({1\over\Gamma_c^2}-{1\over\Gamma_s^2}\right){dM_\pm\over d\Omega}.
\ee
The dynamics of the main shell is now modified with the final term in Equation (\ref{eq:netforce}) being
removed.  

The contact shell attains an equilibrium Lorentz factor given approximately by balancing the first and
second terms on the right-hand side of Equation (\ref{eq:netforcec}),
\be\label{eq:gamceq}
\Gamma_c \simeq  \left[{4\pi (d{\cal E}/d\Omega) {\cal R}\over 8 (R_s\Delta R_s)^2 B_{\rm ex\perp}^2}\right]^{1/4}.
\ee
The prompt phase of the shell deceleration is completed when 
\be
\int (1-\beta_s) dR_s = {R_s\over 4\Gamma_s^2} \simeq {\cal R}.
\ee
This gives 
\ba\label{eq:rdecB2}
R_{{\rm dec},B} &=& \left({2{\cal E} {\cal R}\over B_{\rm ex\perp}^2}\right)^{1/4}\nn
   &=& 1.2\times 10^{13}\;{\cal E}_{41}^{1/4} {\cal R}_{-1}^{1/4} B_{\rm ex\perp,-6}^{-1/2}\quad{\rm cm}.
\ea
This expression is valid as long as $R_{{\rm dec},B} < 2\Gamma_{\rm max}^2{\cal R}$, that is, if
the ambient magnetic field is stronger than the bound (\ref{eq:Bmax}); otherwise Equation (\ref{eq:rdecB})
holds.

Comparing $\Gamma_c$ with the shell Lorentz factor inside the deceleration length (\ref{eq:rdecB2}), we find
\be\label{eq:gamc}
{\Gamma_c\over \Gamma_s} = \left({R_{{\rm dec},B}\over 4\Gamma_s^2 {\cal R}}\right)^{1/2}
\left({R_{{\rm dec},B}\over R_s}\right)^{1/2}.
\ee
At this stage, most of the shell is still accelerating due to the internal Lorentz force and
$\Gamma_s$ is given by Equation (\ref{eq:gammhd}), implying that the right-hand side of Equation (\ref{eq:gamc})
is below unity.

The crossover between deceleration dominated by electron drag and by magnetic drag (corresponding to
$R_{{\rm dec},B} = R_{{\rm dec},e}$) takes place when the ambient Alfv\'en speed reaches
\ba\label{eq:alfcrit}
V_{A\perp} &\equiv& {B_{\rm ex\perp}\over (4\pi n_{\rm ex} m_p)^{1/2}}\nn
&=& 53\,n_{\rm ex,-1}^{5/2} \varepsilon_{\rm th,-1}^{-3/2}{\cal E}_{41} {\cal R}_{-1}^5
\quad {\rm km~s^{-1}}.\nn
\ea
This shows that, if the electromagnetic outflow is warm (e.g. $\varepsilon_{\rm th} \gtrsim 0.01$-0.1), 
then particle drag dominates in strongly ionized parts of the interstellar medium
($n_{\rm ex} \gtrsim 0.1\,\varepsilon_{\rm th,-1}^{3/5}$ cm$^{-3}$), whereas magnetic drag dominates 
in the weakly ionized parts.  The critical ambient electron density
for strong particle drag is reduced in the case of a cool outflow ($\varepsilon_{\rm th} \ll 1$).

\begin{figure}
\epsscale{1.1}
\plotone{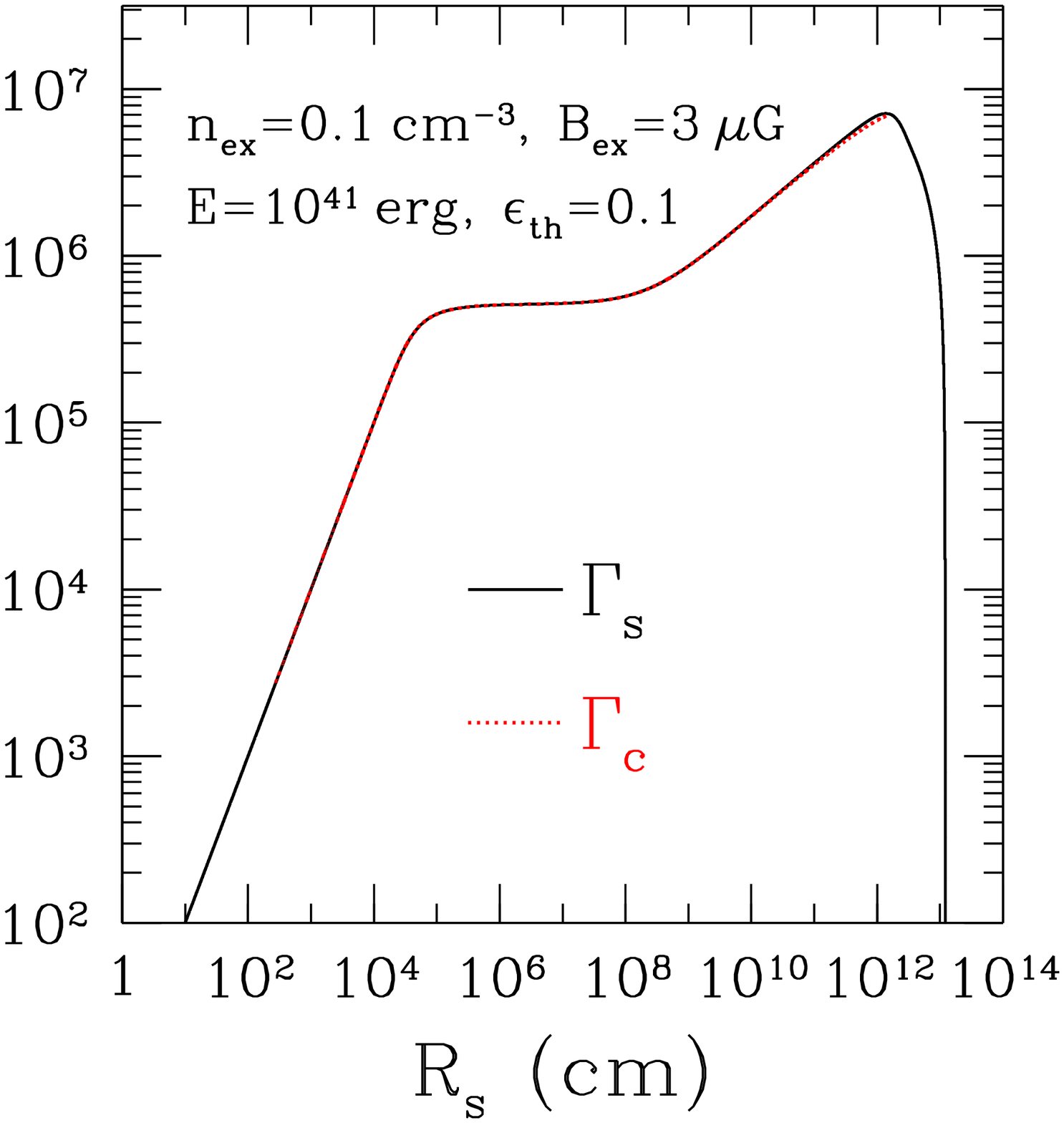}
\plotone{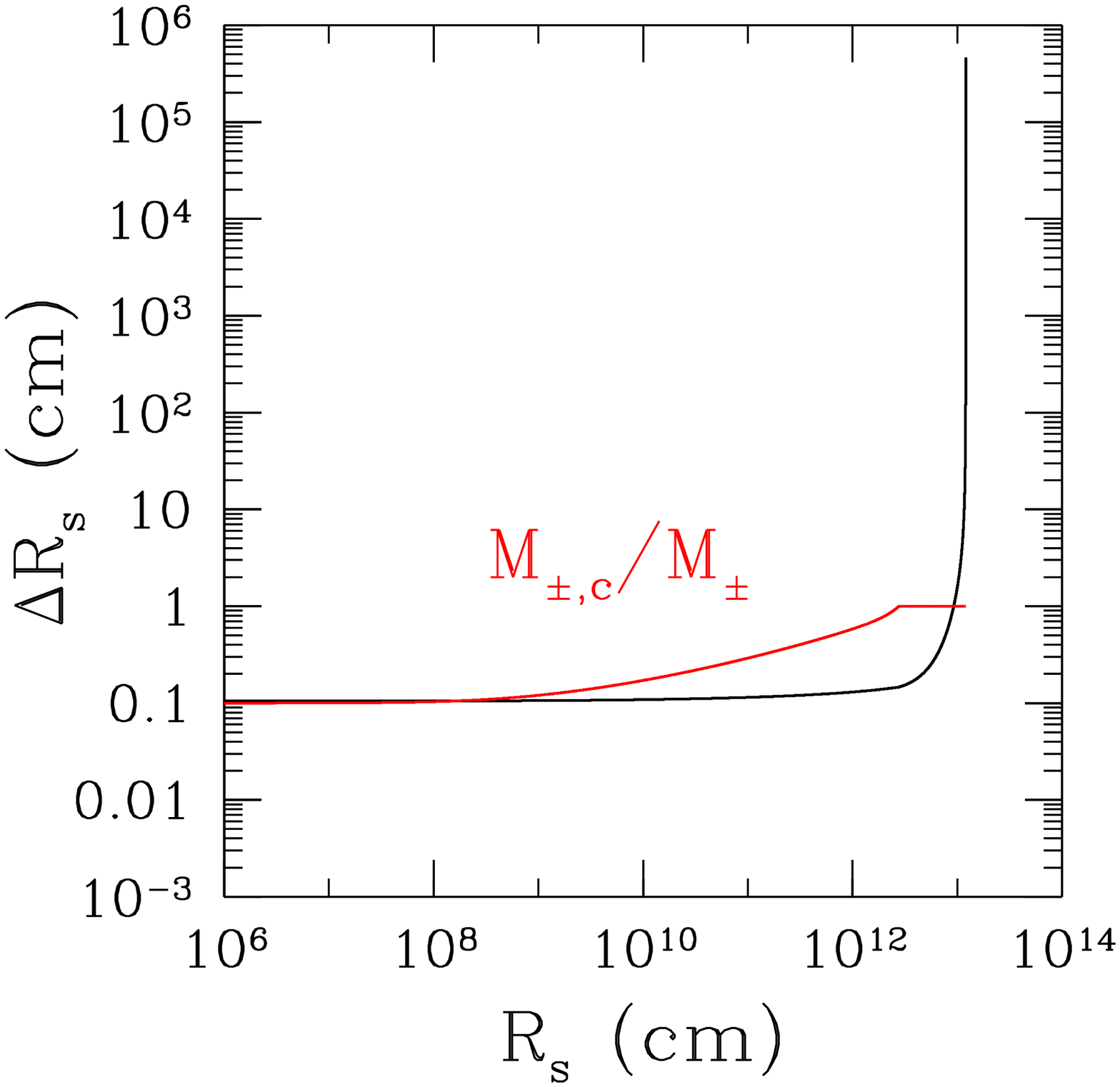}
\vskip .1in
\caption{Warm magnetized shell first accelerates due to anisotropic photon pressure
and then decelerates in a medium with free electron density $n_{\rm ex} = 0.1$ cm$^{-3}$
and magnetic field $B_{\rm ex} = 3\;\mu$G.  Explosion energy $10^{41}$ erg, initial size
${\cal R} = 10^{-1}$ cm, and thermal fraction $\varepsilon_{\rm th} = 0.1$.  
{\it Top panel:}  Mean shell Lorentz factor 
(black curve) and Lorentz factor of forward layer in pressure equilibrium with swept
up external magnetic field (red curve).  Initial linear shell acceleration is
followed by slower $\Gamma_s \sim R_s^{1/3}$ growth due to internal spreading.  The rapid drop begins where drag off the
ambient magnetic field becomes significant ($\Gamma_s  \propto R_s^{-3}$), which becomes
steeper yet as electron drag takes over.  In this case, the shell does not quite reach
the maximum Lorentz factor allowed by internal spreading (Equation \ref{eq:gamax}) before
it decelerates.  {\it Bottom panel:}  Shell thickness versus radius (black curve)
and fraction of shell inertia (in frozen out pairs) that has accumulated in 
the forward contact layer (red curve).}
\vskip .2in
\label{fig:shellkin}
\end{figure}

\begin{figure}
\epsscale{1.1}
\plotone{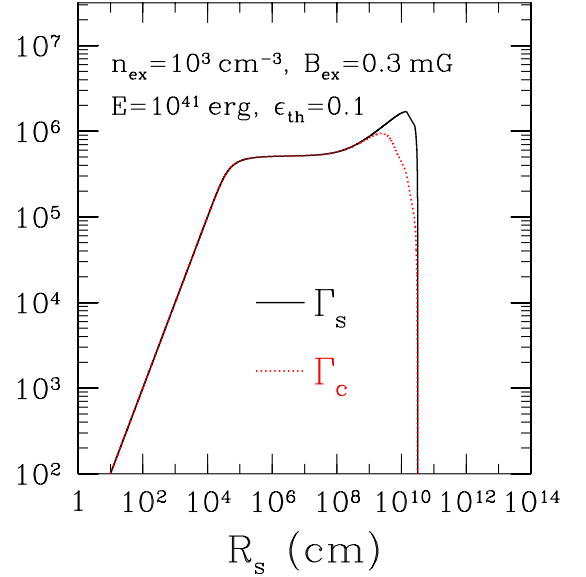}
\vskip .1in
\caption{Same as top panel of Figure \ref{fig:shellkin}, but now for propagation in
a much denser ionized medium with the same ambient Alfv\'en speed.  Now the forward
contact layer (red curve) has a noticeably lower Lorentz factor near the radius of
peak Lorentz factor.}
\vskip .2in
\label{fig:shellkin2}
\end{figure}

\subsection{Sample Shell Trajectories}

The growth of the Lorentz factor and thickness of the magnetized shell is shown in Figures
\ref{fig:shellkin} and \ref{fig:shellkin2} for ISM-like conditions ($n_{\rm ex} \sim 0.1$ cm$^{-3}$,
$B_{\rm ex\perp} = 3\,\mu$G) and much denser ambient plasma ($n_{\rm ex} \sim 10^3$ cm$^{-3}$,
$B_{\rm ex\perp} = 300\,\mu$G).  We evolve a two-layered shell with a forward contact layer in 
approximate pressure equilibrium with the swept up ambient magnetic field, using Equations
(\ref{eq:netforce}), (\ref{eq:drshell}), (\ref{eq:netforcec}), and (\ref{eq:dmcdt}).

An initial phase of linear growth of $\Gamma_s$ is followed by a plateau
after the comoving thermal compactness drops below unity.  Radial spreading of the magnetic
field allows a relatively slow final increase, $\Gamma_s \propto R_s^{1/3}$.  The rapid
deceleration is initiated by the drag from the reflection of the ambient magnetic field,
but quickly the drag from the penetrating ambient charged particles takes over.  In the first
case of an ISM-like medium, the Lorentz factor of the forward contact layer hardly differs from that of the main shell,
which can in practice be treated as a single unit; but in the case of the denser medium, the 
equilibrium Lorentz factor (\ref{eq:gamceq}) lies a factor 2-5 below the Lorentz factor of the main shell.
The variation of shell thickness is shown in the bottom panel of Figure \ref{fig:shellkin}.  
It increases dramatically during the final deceleration according to Equation (\ref{eq:drshell}).
An analytic description of this rapid shell spreading is given in Appendix \ref{a:asymp}.

\subsection{Breakdown of the MHD Approximation}\label{s:mhd}

The magnetic field decouples from the embedded pairs when the comoving plasma frequency drops below
$\sim c/\Delta R_s'$.  The ensuing conversion to a propagating superluminal wave is examined in Section
\ref{s:em}.  

The radius at which this transition takes place is mildly sensitive to details of the heating
of the pairs.   Here the transition radius is estimated by assuming that the random kinetic energy
of the pairs reaches equipartition with the magnetic field in the comoving frame,
\be
n'_\pm \gamma_\pm' m_ec^2  \sim {(B_\perp')^2\over 8\pi} \sim {{\cal E}\over 8\pi\Gamma_s^2 R_s^2\Delta R_s}.
\ee
Combining this with $n_\pm' m_ec^2 \cdot 4\pi R_s^2 \Delta R_s' = {\cal E}/\Gamma_{\rm max}$, we find
\be
{\omega_{P\pm}' \Delta R_s'\over c} = \left({r\over R_{\rm mhd}}\right)^{-1},
\ee
where
\ba\label{eq:rmhd}
R_{\rm mhd} &\;\equiv\;& {\Gamma_s\over\Gamma_{\rm max}} {e\over m_ec^2}\left(2{\cal E} \Delta R_s\right)^{1/2}\nn
            &=& 8.4\times 10^{16}{\Gamma_s\over\Gamma_{\rm max}} {\cal E}_{41}^{1/2} \Delta R_{s,-1}^{1/2}\quad {\rm cm.}
\ea
In the case where the shell is still being accelerated by the internal Lorentz force, one can substitute
Equation (\ref{eq:gammhd}) for $\Gamma_s$ and derive
\be
R_{\rm mhd} = 6.8\times 10^{17}\,\varepsilon_{\rm th,-1}^{3/4}{\cal E}_{41}^{1/2}\quad{\rm cm}.
\ee

The magnetic field is able to escape the shell before it decelerates through its interaction with an 
external magnetic field, but only if the shell starts out with a relatively small thermal energy.
Comparing $R_{\rm mhd}$ with the magnetic deceleration length (\ref{eq:rdecB2}), one obtains the upper bound
\be
\varepsilon_{\rm th} < 4.5\times 10^{-7} {\cal E}_{41}^{-1/3} {\cal R}_{-1}^{1/3} B_{\rm ex\perp,-6}^{-2/3}.
\ee
There is an additional constraint on the ambient electron density:  comparing the deceleration radius $R_{{\rm dec},e}$
as given by Equation (\ref{eq:rdece}) with $R_{\rm mhd}$, one sees that
\be
n_{\rm ex} < 3\times 10^{-5}  {\cal E}_{41}^{-1/2} {\cal R}_{-1}^{-3/2}\quad{\rm cm}^{-3}.
\ee
Direct escape of the embedded magnetic field is therefore most relevant for tiny explosions of a 
low thermal energy and short duration, ${\cal R} < 1$ mm.

\section{Electromagnetic Emission}\label{s:radio}

The low-frequency electromagnetic emission from a thin and ultra-relativistic shell has three major
contributions:
\vskip .05in
\noindent 1. Reflection of an ambient static magnetic field into a propagating transverse wave
\citep{rees77,blandford77}.
\vskip .05in
\noindent 2. Direct linear conversion of an embedded magnetic field into a propagating superluminal wave. 
\vskip .05in
\noindent 3.~Excitation of an electromagnetic mode outside the shell by a corrugation of its surface.
This external mode can either escape directly or tunnel away from the shell as it decelerates, depending
on whether the phase speed of the corrugation is larger than or small than the speed of light.
\vskip .05in
We consider each of these channels in turn.

\subsection{Reflection of Ambient Magnetic Field}\label{s:reflect}

Consider the transverse vector potential that is excited at radius $r$ and time $t > r/c > R_{\rm sh}(t)/c$ by the
outward motion of a spherical conducting shell:
\be
r {\bf A}_\perp(r,t) = 4\pi\int dt' dr' G(r-r',t-t') r'{\bf J}_\perp(r',t').
\ee
In the case of one-sided emission, the Green function is $G(r-r',t-t') = 1$ for $r-r' < c(t-t')$ and 0 otherwise.
The surface current
\be
{\bf K}_\perp(t') = \int dr' {\bf J}_\perp(r',t') = {\dot R_s(t')\over 4\pi}\,\hat r\times {\bf B}_{\rm ex}.
\ee
Then at a radius $r \gg {\cal R}$,
\be
r {\bf A}_\perp(r,t) = {1\over 2} R_s^2(t_{\rm em})\, \hat r\times {\bf B}_{\rm ex}.
\ee
Here $t_{\rm em} < t$ is the maximum time at which an electromagnetic pulse leaving the emitting surface
of the shell can reach radius $r$:
\be
r - R_s(t_{\rm em}) = c(t - t_{\rm em}).
\ee
The electric field is
\be
r{\bf E}_\perp(r,t) = -{1\over c}{\partial {\bf A}_\perp\over\partial t} 
\simeq -2\Gamma_s^2(t_{\rm em}) R_s(t_{\rm em})\,\hat r\times {\bf B}_{\rm ex},
\ee
where we have made use of
\be
{\partial t_{\rm em}(r,t)\over\partial t} \simeq 2\Gamma_s^2(t_{\rm em}).
\ee

\begin{figure}
\epsscale{1.1}
\plotone{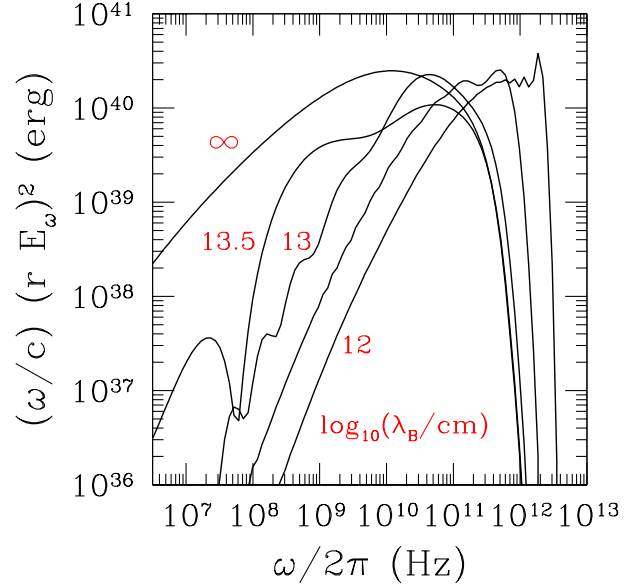}
\vskip .1in
\caption{Spectrum of the electromagnetic pulse produced by reflection of the ambient
magnetic field, in the shell trajectory shown in Figure \ref{fig:shellkin}.  The
spectrum below the peak corresponds to the deceleration phase,
and the steep high-frequency part to the acceleration phase.  In addition to shell
propagation through a uniform magnetic field, we show the effect of adding a harmonic modulation
of the sign of the external magnetic field, $B_{\rm ex\perp}(R_s) \propto \cos(2\pi R_s/\lambda_B)$.
Now sharp spectral features emerge, which represent the Lorentz upscattering of the spatial
structure in the ambient magnetic field.}
\vskip .2in
\label{fig:spectrum}
\end{figure}

The energy radiated in a given spectral band is obtained by integrating at some large radius
placed outside the shell's deceleration volume,
\ba
&&4\pi {d(\omega {\cal E}_\omega)\over d\Omega} = r^2 \biggl| \int_0^\infty cd(t-r/c) 
      e^{-i\omega(t-r/c)} E_\perp^2(r,t)\biggr|\nn
&&= 2\biggl|\int_0^\infty cdt_{\rm em} e^{-i\omega[t_{\rm em}-R_s(t_{\rm em})]}
   [B_{\rm ex\perp}R_s]^2(t_{\rm em}) \Gamma^2(t_{\rm em})\biggr|.\nn
\ea
The total energy radiated is
\be
4\pi{d\over d\Omega}\int d\omega {\cal E}_\omega = 2\int_0^\infty cdt_{\rm em}
[B_{\rm ex\perp}R_s]^2(t_{\rm em}) \Gamma^2(t_{\rm em}).
\ee

Figure \ref{fig:spectrum} shows the spectral distribution of the reflected superluminal mode produced by the
shell trajectory of Figure \ref{fig:shellkin}, assuming a homogeneous ambient magnetic field.  The peak
of the spectrum is quite broad in the case of a homogeneous ambient magnetic field.  Adding a periodic
reversal of this field, $B_{\rm ex\perp} \propto \cos(2\pi R_s/\lambda_B)$, where $\lambda_B$ is
a variable wavelength, produces significant spectral
structure on the low-frequency tail, and if $\lambda_B$ is small enough also produces a high-frequency
extension of the spectrum.  Related sharp spectral features appear in the bursts of the repeating source
FRB 121102 \citep{scholz16}.

\subsection{Linear Conversion to a Superluminal Wave}\label{s:em}

Here we consider the expansion of a finite slab of magnetic field with embedded electron-positron pairs. 
The spherical shell problem is mapped onto a cartesian geometry by including expansion parallel to the
shell.   We generalize the MHD problem examined by \cite{granot11} and \cite{lyutikov10}
by separating the dynamics of the embedded charges from the time evolution of the electromagnetic field.
The initial configuration is magnetically dominated, but as $\omega_{P\pm}'$ drops below $\sim c/\Delta R_s'$
a superluminal mode with $|{\bf E}| > |{\bf B}|$ emerges.  An analytical approach to the related problem
of strong wave dissipation in pulsar winds can be found in \cite{mm96}.

\begin{figure}
\epsscale{1}
\plotone{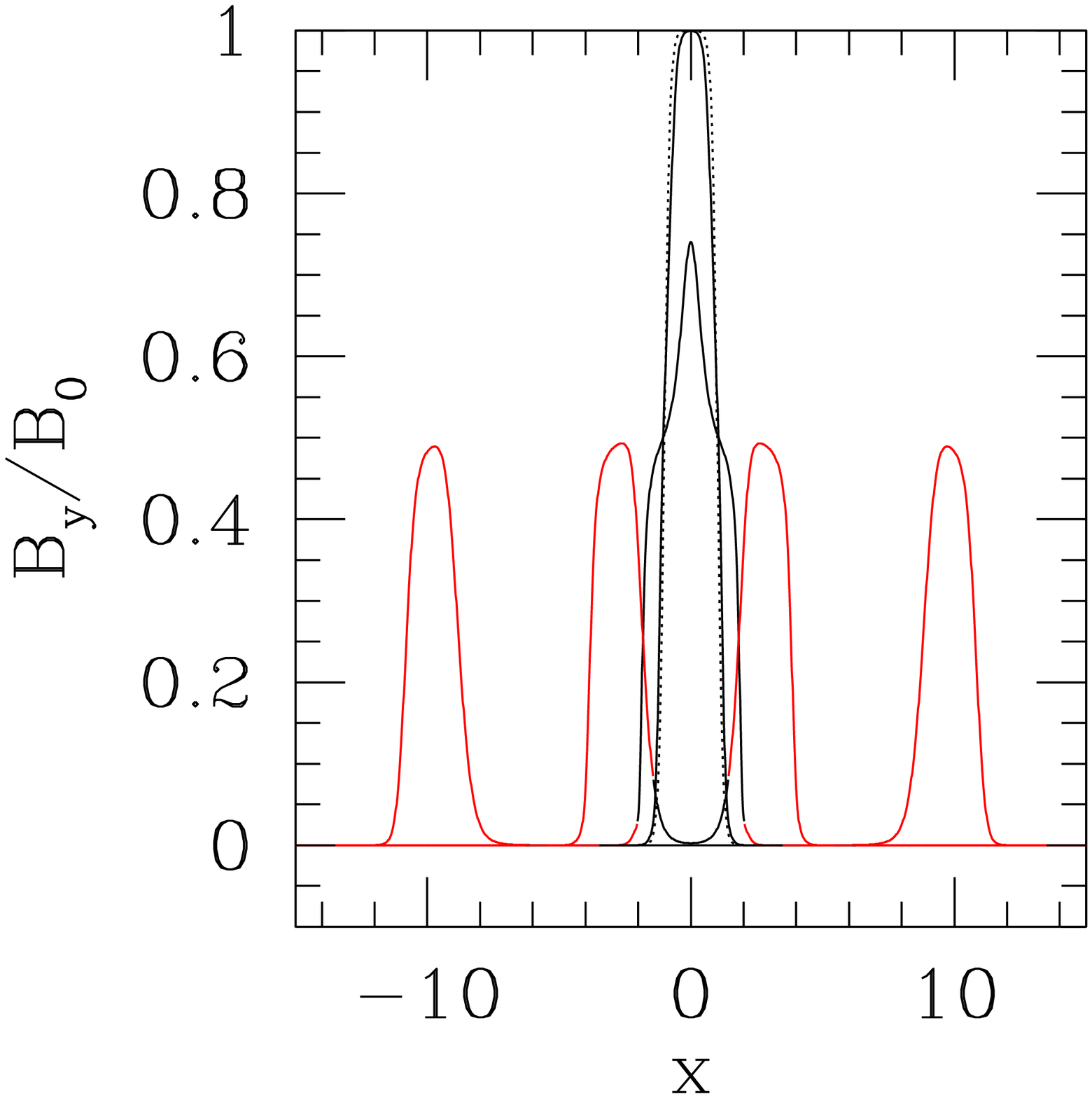}
\plotone{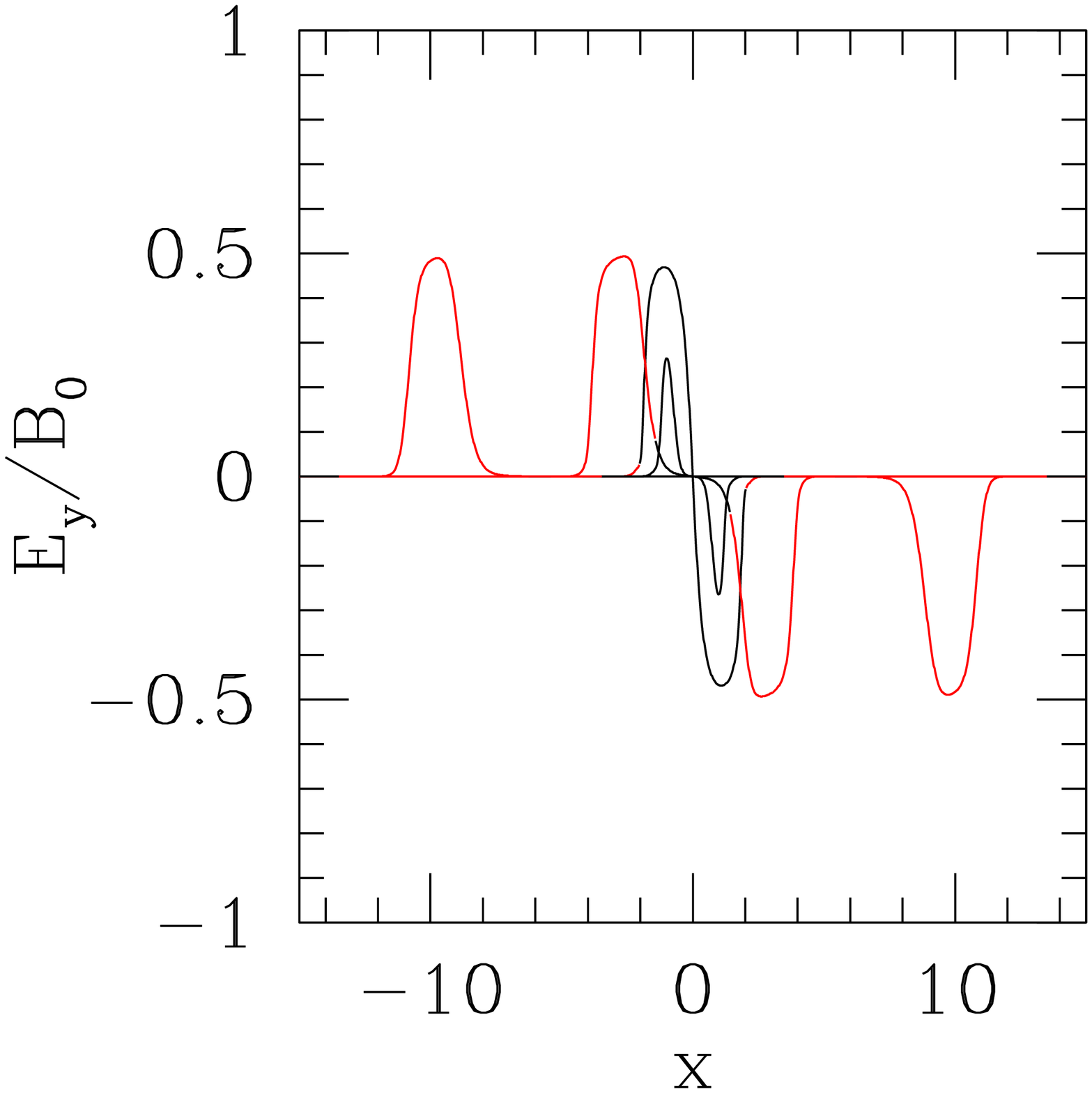}
\vskip .1in
\caption{Magnetic and electric fields emerging from a magnetized shell with normalized initial configuration
$B_y(x) = n_\pm(x) = (1/2)[\tanh(x-L) - \tanh(x+L)]$ with $L=1$;  $E_z(x) = 0$; $v_z(x)/c = B_y'(x)/4\pi en_\pm$;
and $\sigma_\pm(0) = 1$, $\omega_{P\pm}(0) = 100c/L$.  Shell expands on a timescale $t_{\rm exp} = 0.2L/c$
(Equation (\ref{eq:eom})).  Dotted curve shows initial magnetic field.  Solid curves show snapshots at
times 0.3, 1, 3, 10 $L/c$, with the $x$-extension of the waveform increasing with time.
Black (red) colors correspond to zones where $(B_y^2 -E_z^2)/(B_y^2+E_z^2) > (<) 10^{-3}$.}
\vskip .2in
\label{fig:fields}
\end{figure}

\begin{figure}
\epsscale{1}
\plotone{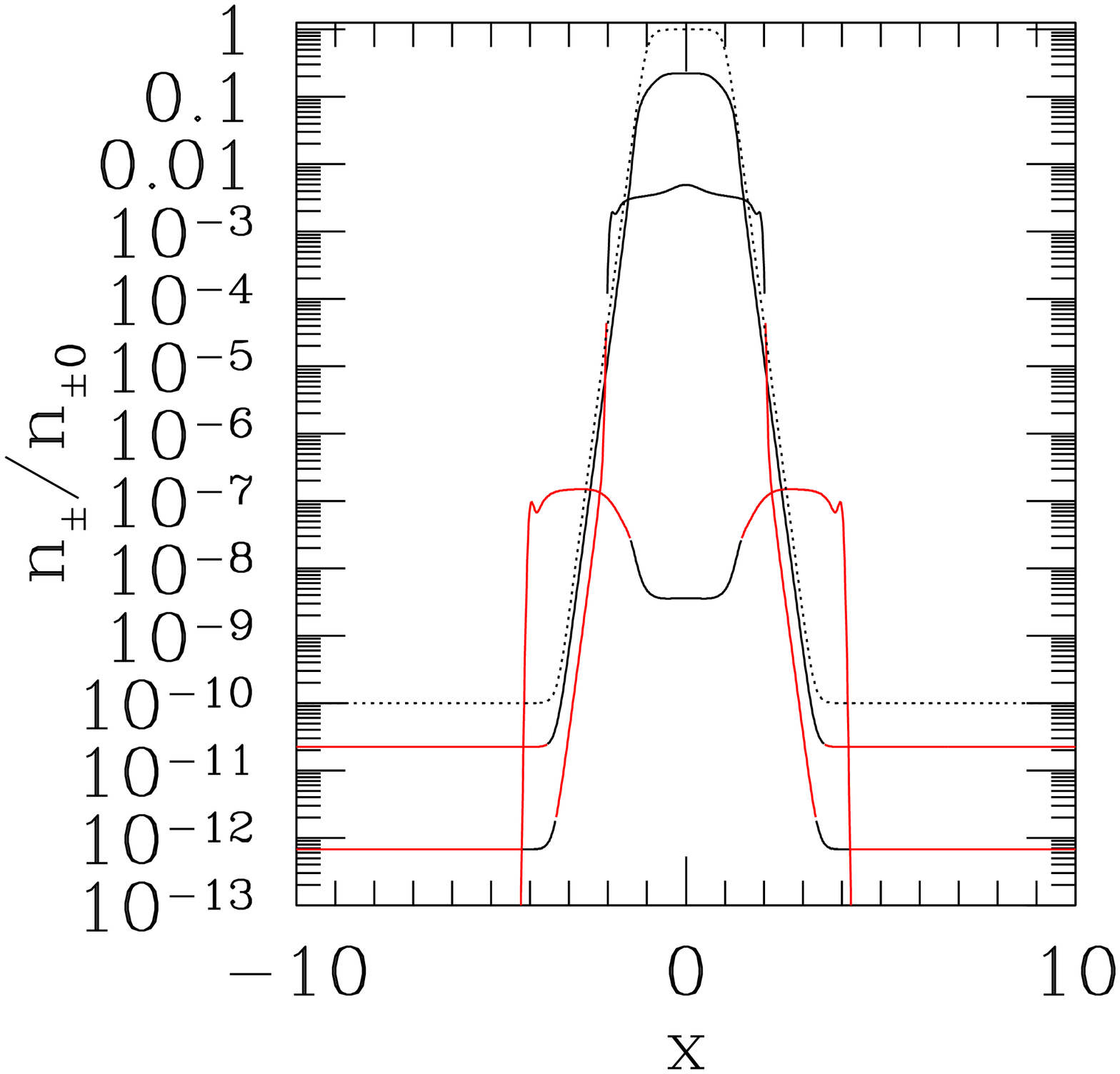}
\plotone{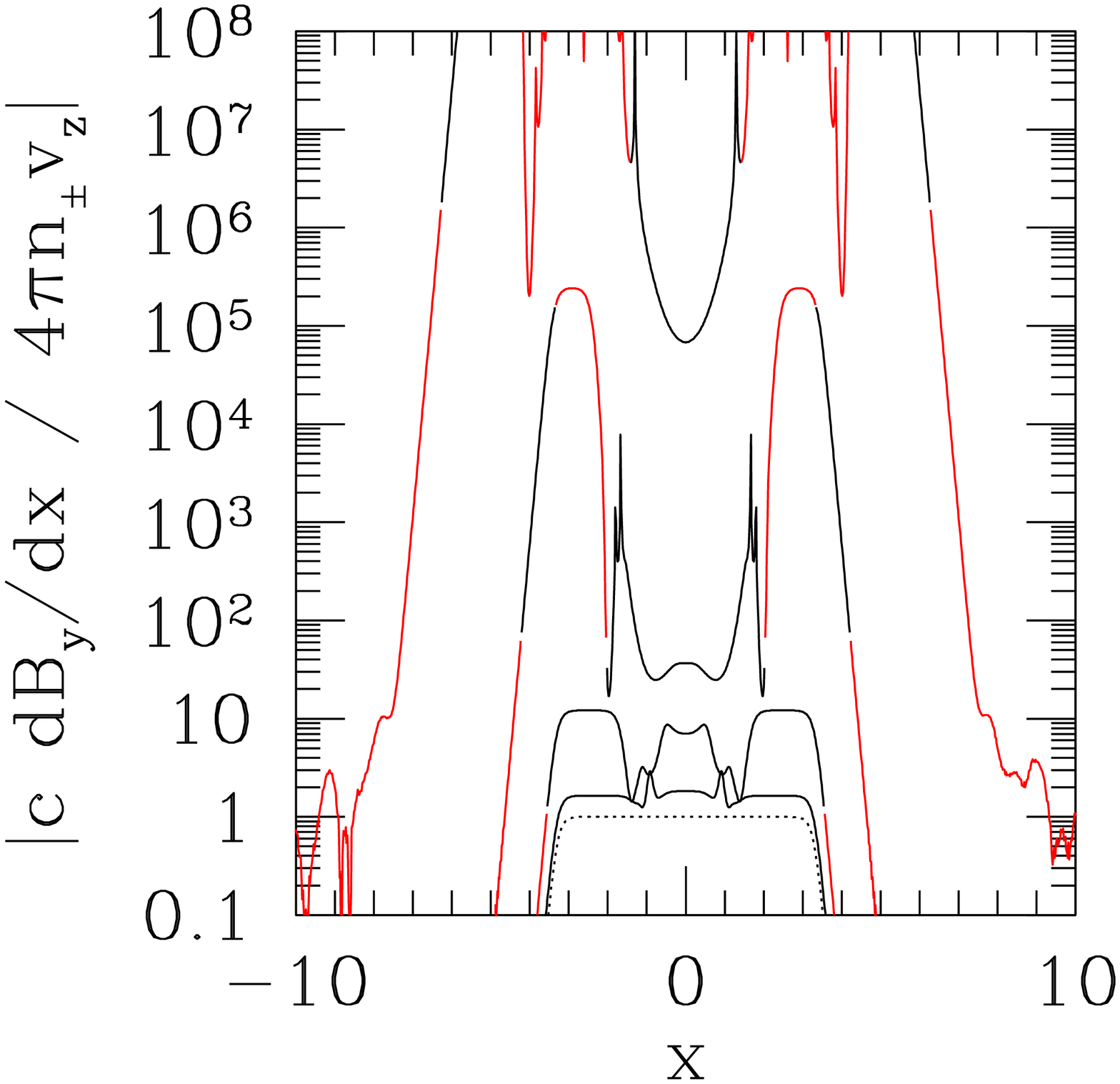}
\vskip .1in
\caption{{\it Top panel:} 
Density $n_\pm$ of electrons and positrons relative to initial density (dotted), at times
0.3, 1, 3, $c/L$, with density at $x=0$ decreasing with time.
Black (red) colors correspond to zones where $(B_y^2 -E_z^2)/(B_y^2+E_z^2) > (<) 10^{-3}$.
{\it Bottom panel:}  Ability of pair plasma to suppress the displacement current at times
0.1, 0.3, 1, 3 $c/L$, with value at $x=0$ increasing with time. The top curve
extends beyond the plotted range, to $\sim 10^{13}$.}
\vskip .2in
\label{fig:cs}
\end{figure}

\begin{figure}
\epsscale{1}
\plotone{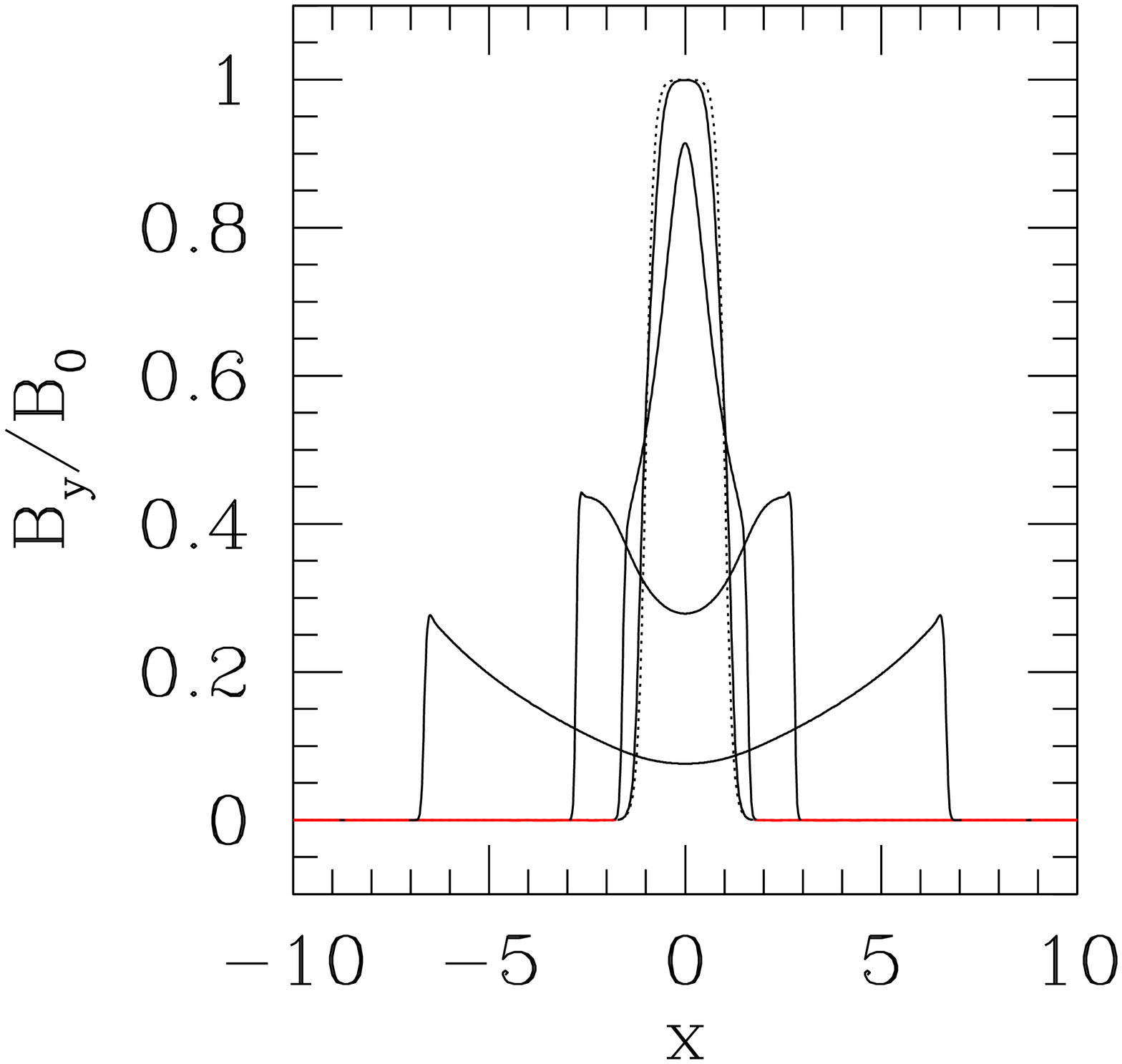}
\plotone{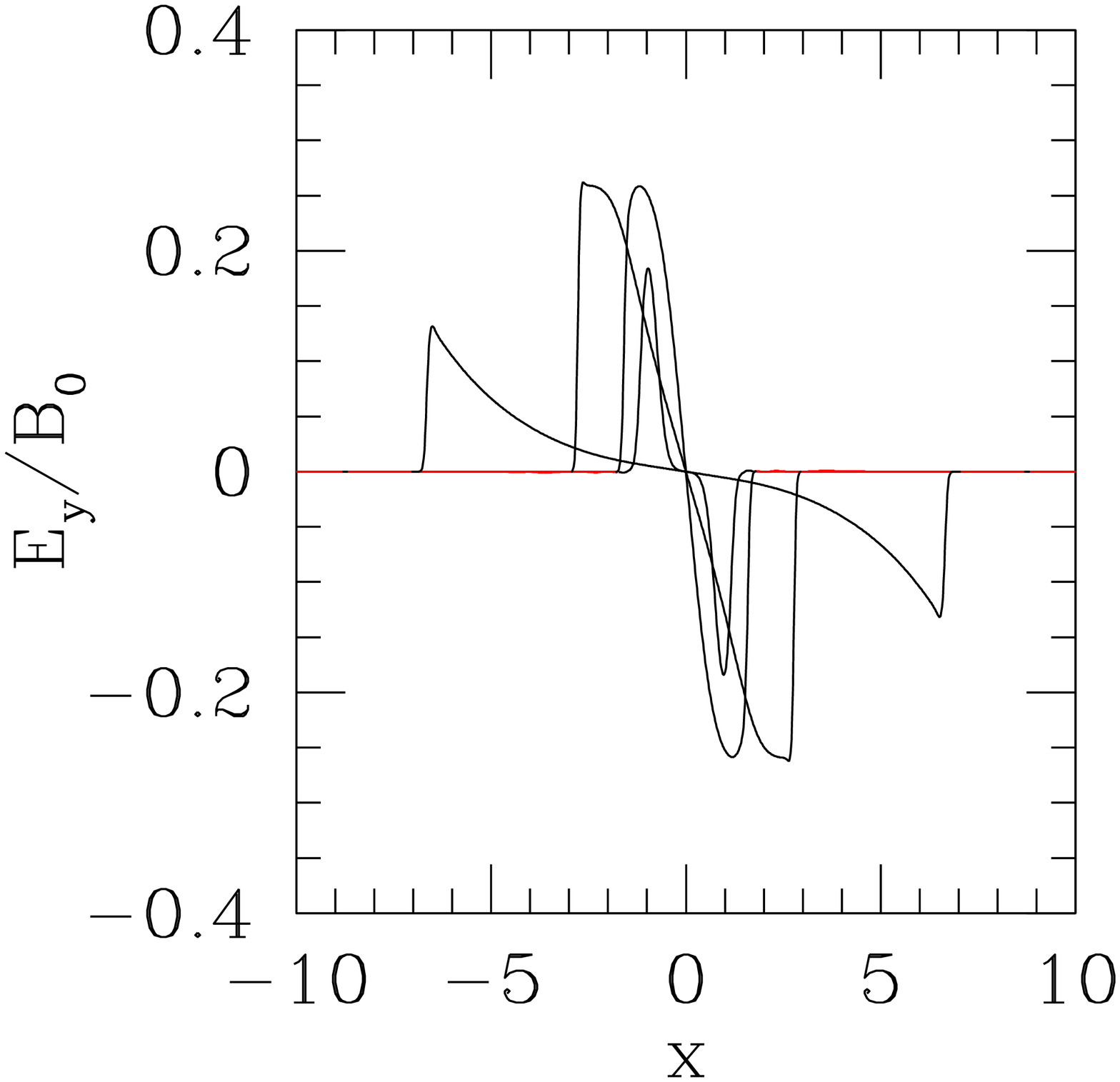}
\plotone{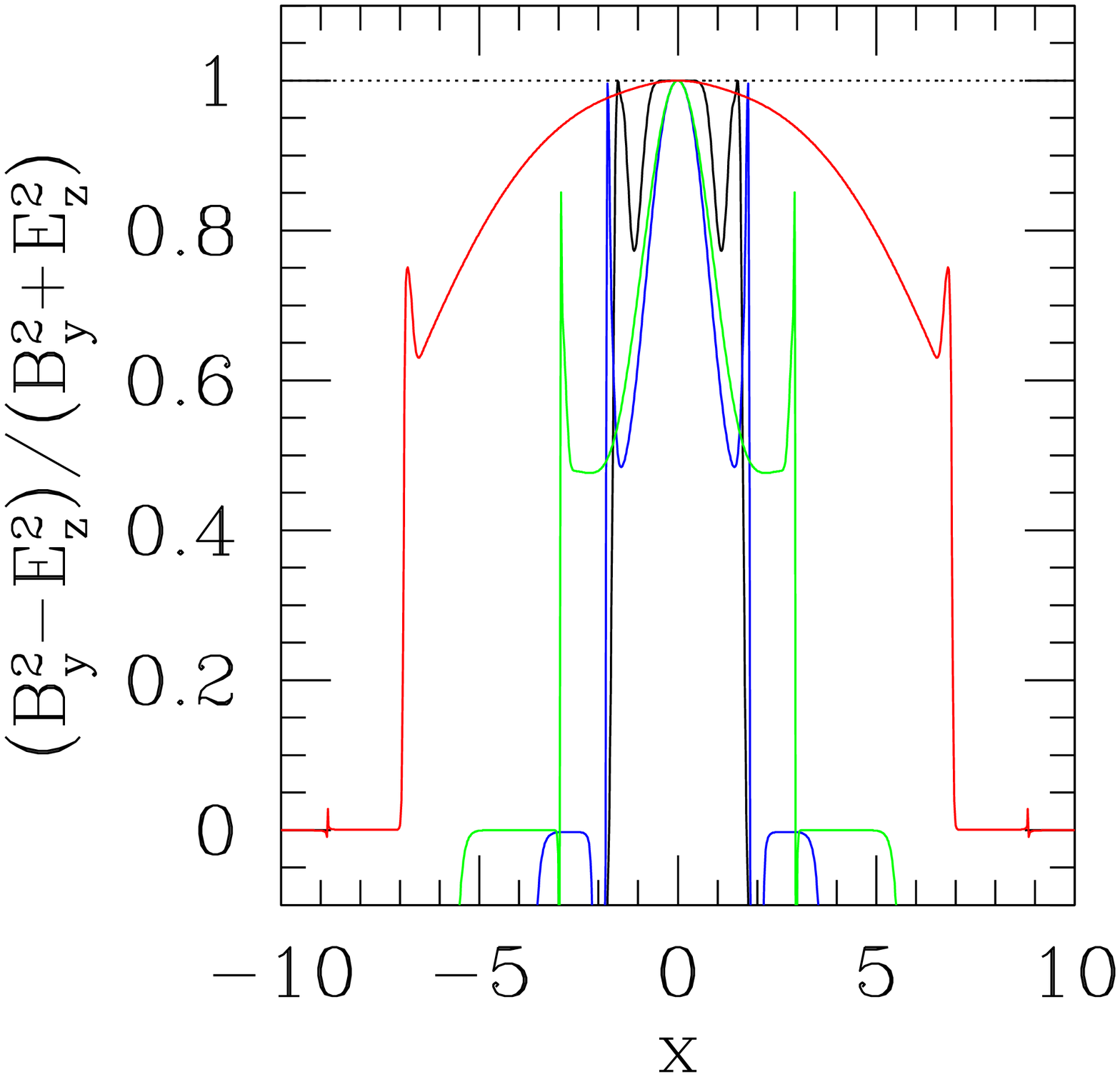}
\vskip .1in
\caption{Comparison with Figure \ref{fig:fields}, with same initial conditions but now $t_{\rm exp} = \infty$.
Maximum $x$-extension of the pulse increases with time.  The curves are colored in the bottom panel
so as to more clearly distinguish them.}
\label{fig:fields2}
\vskip .2in
\end{figure}

\begin{figure}
\epsscale{1}
\plotone{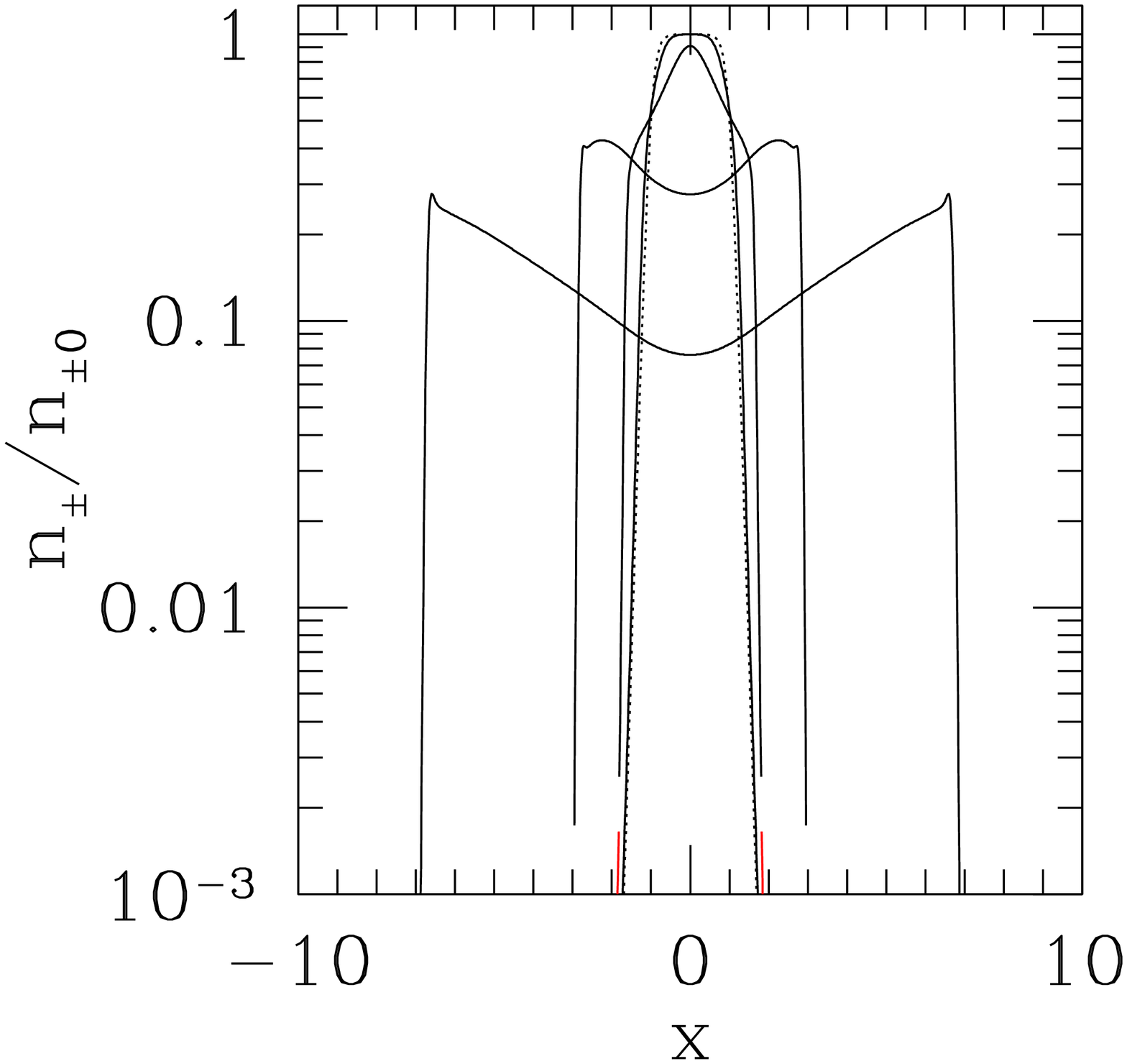}
\vskip .1in
\caption{Comparison with top panel of Figure \ref{fig:cs}, with same initial conditions but now $t_{\rm exp} = \infty$.}
\label{fig:cs2}
\vskip .2in
\end{figure}

We work in the comoving frame as defined by vanishing radial particle velocity at the mid-point of the 
shell.\footnote{We drop all primes in the remainder of this section for notational simplicity.}
The radial coordinate is mapped onto $r \rightarrow x$, and the magnetic field is locally
straight, ${\bf B} = B_y(x) \hat y$ (Figure \ref{fig:fields}).   The embedded cold 
electrons and positrons have total density $n_\pm(x) = n_+(x) + n_-(x)$, vanishing net charge density, and 
associated plasma frequency $\omega_\pm = (4\pi n_\pm e^2/m_e)^{1/2}$.  For simplicity, the electric field is
assumed to vanish initially, and the positive and negative charges to drift oppositely in the $z$-direction,
thereby supplying a conduction current 
\be\label{eq:jz}
J_z = e (n_+v_{z+} - n_-v_{z-}) = e n_\pm v_{z+} = {c\over 4\pi}\partial_xB_y.
\ee
The Lorentz force drives a particle drift in the $x$ direction with $v_{x+} = v_{x-} = v_{x\pm}$.  This drift can only be 
viewed as a hydromagnetic displacement, $v_x = -(E_z/B_y)c$, when the plasma is relatively dense, corresponding to 
$\omega_{P\pm} \Delta R_s/c \equiv (\omega_{P\pm}/c) |B_y|/|\partial_x B_y| \gg 1$.  

To understand the basic behavior, consider the kinetic energy that must be stored in the $z$-drift of the
pairs in order to inhibit the growth of the electric field.   We are interested in a late phase of the shell
expansion when $\sigma'_\pm \ll 1$, and so the drift is subrelativistic even in the outer parts of the shell
where one may find $|{\bf E}| > |{\bf B}|$.  Therefore
\be
n_\pm {1\over 2}m_e v_{+z}^2 \sim \left({\omega_{P\pm} \Delta R_s\over c}\right)^{-2} {B_y^2\over 8\pi}.
\ee
When $\omega_{P\pm} \Delta R_s/c \gg 1$, the particles need tap only a small fraction of the magnetic field energy.
The magnitude of the $x$-drift is found from 
\be 
m_e {v_{x \pm}^2\over \Delta R_s} \sim {ev_{+z} B_y\over c},
\ee
corresponding to expansion at the fast mode speed, $v_{x\pm} \sim v_F = B_y/(4\pi n_\pm m_e)^{1/2}$.   The 
radial particle drift relaxes to equilibrium on the timescale 
$(\omega_{P\pm} \Delta R_s/c)^{-1} (\Delta R_s/v_{x\pm})$.  

The shell during its expansion makes a transition from $\omega_{P\pm} \Delta R_s/c \gg 1$ to $\omega_{P\pm} \Delta R_s/c \ll 1$,
since
\be
\left({\omega_{P\pm} \Delta R_s\over c}\right)^2 \propto n_\pm (\Delta R_s)^2 \propto M_\pm {\Delta R_s \over R_s^2}
\propto M_\pm R_s^{-4/3}.  
\ee
Radial spreading of the shell dilutes $n_\pm$ and $B$ in equal proportions, as does 
expansion in the direct normal to ${\bf r}$ and to ${\bf B}$.  However non-radial expansion along ${\bf B}$ only
dilutes the particles;  this can be taken into account by adding a damping term to the evolution equation
for $n_\pm$ with characteristic timescale $t_{\rm exp} \sim \Delta R_s/v_F$.  In this way we can study the transition
away from a hydromagnetic state, starting with only a fraction of the energy in particle drift ($\omega_{P\pm} \Delta R_s/c \ll 1$).

The equations to be solved are therefore
\ba\label{eq:eom}
{\partial B_y\over\partial t} &=& c{\partial E_z\over\partial x};\nn
{\partial E_y\over\partial t} &=& -4\pi e n_\pm v_{z+} + c{\partial B_y\over\partial x};\nn
{dv_{z+}\over dt} &=& {\partial v_{z+}\over\partial t} + v_{x\pm}{\partial v_{z+}\over\partial x} = 
       {e\over m_e}\left(E_z + {v_{x\pm}\over c}B_y\right);\nn
{dv_{x\pm}\over dt} &=& {\partial v_{x\pm}\over\partial t} + v_{x\pm}{\partial v_{x\pm}\over\partial x} = 
-{v_{z+}\over m_ec}B_y;\nn
{\partial n_\pm\over \partial t} &=& -{\partial(n_\pm v_{x\pm})\over\partial x} - {n_\pm \over t_{\rm exp}}.\nn
\ea
A numerical solution is easily found using a one-dimensional spectral code.  

One integral is immediately obtained from the equation of motion for $v_{z+}$, which expresses the conservation
of the generalized momentum of the particles parallel to the shell,
\be
v_{z+} - v_{z+}(0) = {e\over m_e c}\left[A_z - A_z(0)\right] = {e\over m_e c}\left[\Phi_B - \Phi_B(0)\right].
\ee
The vector potential is related to the non-radial flux by $A_z = \Phi_B \equiv \int^\infty_x B_y dx$.
If an electron or positron starts with a relatively slow drift, it can reach a speed no larger than
$\sim e\Delta A_z/m_ec$, where $\Delta A_z$ is the net flux threading the shell.  Therefore
\be
{v_{+z}\over c} \lesssim \sigma_\pm^{1/2} {\omega_{P\pm}\Delta R_s\over c},
\ee
and relativistic drift cannot develop during the transition to $\omega_{P\pm} \Delta R_s/c \lesssim 1$,
starting from a state with $\sigma_\pm \ll 1$.

Figures \ref{fig:fields}-\ref{fig:cs2}
show the development of a magnetized shell starting with a range of plasma densities.  In the first
two figures, expansion parallel to the magnetic field is included.  Outer zones with 
$|{\bf E}| > |{\bf B}|$ make an appearance as $\omega_{P\pm} \Delta R_s/c$ is reduced,
corresponding to an escaping superluminal wave.  In the second two figures, the expansion parallel
to ${\bf B}$ is not included, and the spreading remains essentially hydromagnetic with $|{\bf E}| < |{\bf B}|$.
Then a caustic-like structure forms at the outer boundary of the expanding wave.

\subsection{Emission due to Shell Corrugation}\label{s:poycorr}

Corrugation of the shell opens up an additional low-frequency emission channel:   the excitation
of an electromagnetic wave with finite wavenumber $k_\parallel$ parallel to the shell surface.
This mode is trapped near the shell if the phase speed is $\omega/k_\parallel < c$, that is,
if the exciting MHD mode within the shell is an Alfv\'en mode.   On the other hand, two
anti-propagating Alfv\'en modes can non-linearly convert to a fast mode which propagates
toward the shell surface.  The resulting surface displacement has a pattern speed 
$\omega/k_\parallel > c$, where now $k_\parallel < |{\bf k}|$ is the projection of the wavevector
onto the shell surface.  

The corrugation is most plausibly excited by reconnection of the shell magnetic field with
the field swept up from the ambient medium.    When this ambient field dominates the drag on 
the shell, reconnection is an efficient mechanism of exciting irregularities.  That is because the
magnetic flux  $\Phi_{\rm ex} = B_{\rm ex} R_s$ swept up is comparable to, or larger than, the flux
in the shell.  Reconnection is facilitated if the two fluxes have opposing signs.  
Equation (\ref{eq:rdecB}) implies that 
\be
{\Phi_{\rm ex} \over B_s \Delta R_s} \sim \left({R_{{\rm dec},B}\over \Gamma_{\rm max}^2 {\cal R}}\right)^{1/2}
\ee
if the deceleration occurs around or outside radius $R_{\rm sat}$ (Equation \ref{eq:rsat}).
Variations in the ambient electron density
or magnetic energy density could be present on a small angular scale $\sim 1/\Gamma_s$,  but
they would then average out (by a factor $\sim \Gamma_s^{-1/2}$) in the direction of propagation 
of the shell.

\begin{figure}
\epsscale{1.1}
\plotone{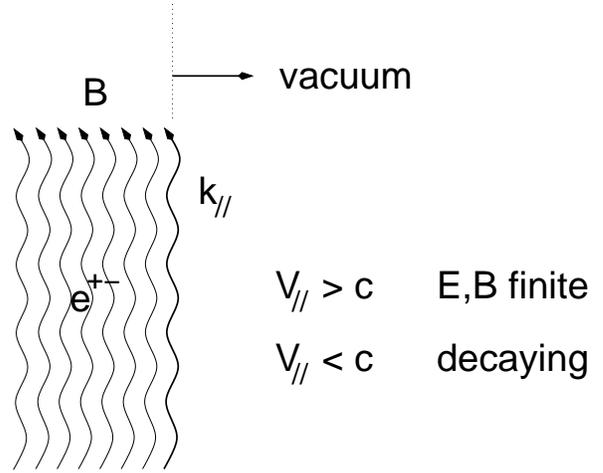}
\vskip .1in
\caption{The surface of a magnetized shell with embedded electrons and positrons is
corrugated by an internal hydromagnetic mode.   In the case of an Alfv\'en-like mode, 
this corrugation has phase speed $V_\parallel < c$ parallel to the shell surface, 
but for an obliquely propagating fast mode $V_\parallel > c$.  The corrugation
combined with the continuity of ${\bf B}_\perp$ and ${\bf E}_\parallel$ implies
the excitation of a vacuum electromagnetic mode outside the shell.  This mode
is trapped, decaying exponentially away from the shell surface, when $V_\parallel < c$,
but transmits finite normal Poynting flux when $V_\parallel > c$.  Even in the first
case, net Poynting flux is transmitted to infinity if the shell surface is
decelerating (Figure \ref{fig:poynting}).   A normal force balance across the shell
surface is achieved in the presence of an external vacuum magnetic field, which is
not plotted for clarity.}
\vskip .2in
\label{fig:corrug}
\end{figure}

\begin{figure}
\epsscale{1.1}
\plotone{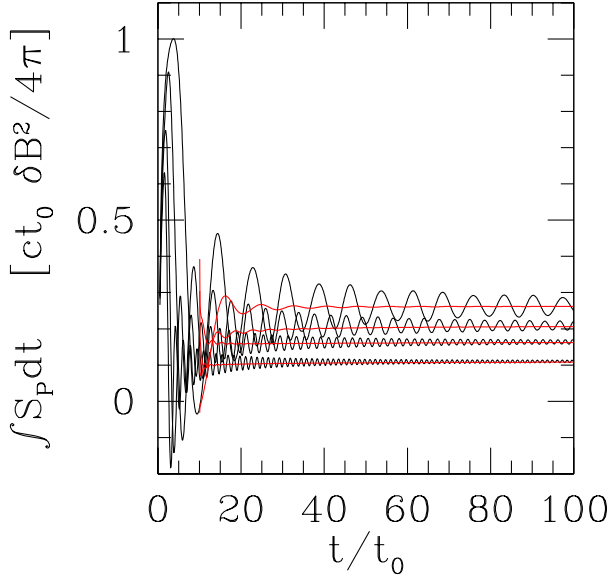}
\vskip .1in
\caption{Black curves: time integral of the Poynting flux emanating from the surface of a magnetized shell
whose surface is corrugated with wavenumber $k_\parallel = \{1,2,4,6\} (V_\parallel t_0)^{-1}$, with
$V_\parallel = c/\sqrt{2}$.  Red curves: running average of the oscillations.  The Lorentz factor of 
the shell scales as $\Gamma_s \propto R_s^{-1}$ in the frame of the explosion.  The calculation is
done is a frame boosted by Lorentz factor $\Gamma_0$, which equals $\Gamma_s$ at time $t_0$ in the
boosted frame.}
\vskip .2in
\label{fig:poynting}
\end{figure}

The shell surface is represented here in cartesian geometry, with the half-space $x < 0$ (labeled
$-$) filled with a perfectly conducting fluid and $x > 0$ (labeled $+$) empty of charged particles
(Figure \ref{fig:corrug}).  The entire space is filled with a uniform background magnetic field $B \hat y$.

The mode excited in at $x < 0$ corrugates the surface of the magnetofluid, producing a magnetic
disturbance
\be
\delta B_x^-(x=0) \;=\; \delta B_x^+(x=0) \;=\; \delta B_0 e^{ik_\parallel(y - V_\parallel t)}.
\ee
Faraday's law implies an electric perturbation
\be
\delta E_z^-(x=0) \;=\; \delta E_z^+(x=0) \;=\; {V_\parallel\over c}\delta B_0 e^{ik_\parallel(y - V_\parallel t)}.
\ee
The electromagnetic perturbation outside the shell is therefore described by the vector potential
\be\label{eq:az}
\delta A_z^+(x>0)\;=\; -i {\delta B_0\over k_\parallel} f(x) e^{ik_\parallel(y - V_\parallel t)}.
\ee
Substituting into the wave equation gives
\be 
f(x) = \left\{                                                                                                                                
     \begin{array}{ll}                                                                                                                             
      e^{-k_A x}  \quad & k_A^2 = k_\parallel^2(1-V_\parallel^2/c^2) \quad\quad (V_\parallel < c) \\
      e^{ik_F x}  \quad & k_F^2 = k_\parallel^2(V_\parallel^2/c^2-1) \quad\quad (V_\parallel > c).
     \end{array}                                                                                                                                   
   \right.                                                                                                                                         
\ee                
The labels $A$ and $F$ refer to the exciting mode being i) an Alfv\'en mode propagating parallel to
the shell surface;  or ii) to a fast mode incident obliquely on the surface from $x < 0$. 

We are so far considering a shell whose mean position does not accelerate with respect to the
observer.  Then the Poynting flux flowing to positive $x$ vanishes in case $A$ but not in case $F$:
\ba
{\bf S}_{\rm P} &=& {\rm Re}\left({\delta E_z\,\delta B_y^*\over 4\pi}c \right) \hat x \nn
                &=& \left\{ 
                    \begin{array}{ll}
               0 \quad\quad &(V_\parallel < c) \\
               (k_F/k_\parallel) (\delta B_0^2/4\pi) c\,\hat x \quad\quad &(V_\parallel > c).
     \end{array}                                                                                                                                   
   \right.                                                                                                                                         
\ea             

In the more realistic case of a decelerating shell, radiation is emitted by both types of corrugation.
We demonstrate this with a concrete example.   The corrugation implies a surface current 
\be
\delta{\bf K}(y,t) = {c\delta B_0\hat z\over 4\pi \Gamma_{\rm sh}(t)}e^{i[k_\parallel y-\phi(t)]},
\ee
as measured in a fixed inertial frame.  Here 
\be
\phi(t) = k_\parallel V_\parallel \int {dt\over \Gamma_{\rm sh}(t)}
\ee
is the phase of the wave in this frame.  The gauge potential outside the shell is
\be
\delta{\bf A}^+(x>x_{\rm sh},y,t) = 4\pi\int dt' \delta{\bf K}(y,t') G^+\bigl[x-x_{\rm sh}(t'),t-t'\bigr],
\ee
where $x_{\rm sh}(t)$ is the position of the shell surface.  The one-sided Green function is
\be
G^+(x-x',t-t') = J_0\left[k_\parallel\sqrt{c^2(t-t')^2-(x-x')^2}\right].
\ee
Hence
\ba
\delta A_z^+(x>x_{\rm sh},t) &=&  \delta B_0 \int {cdt'\over \Gamma_{\rm sh}(t')}\,
\biggl\{ e^{i[k_\parallel y - \phi(t')]} \times \nn
  && \quad J_0\left[k_\parallel\sqrt{c^2(t-t')^2-(x-x_{\rm sh}(t'))^2}\right]\biggr\}.\nn
\ea
The solution (\ref{eq:az}) is readily recovered by setting $\Gamma_{\rm sh} = 1$,
$x_{\rm sh} = 0$.

It is convenient to work in a frame boosted by Lorentz factor $\Gamma_0$ with respect to the
center of the explosion, i.e. corresponding to the peak Lorentz factor of the shell.
If $\Gamma_s$ decreases with radius as $R_s^{-\alpha}$, then 
in the boosted frame $\Gamma_s(t) \propto t^{-\alpha/(1+\alpha)}$, and
the mean surface speed in this frame is
\be
\beta_{\rm sh}(t) = {1 - (t/t_0)^{2\alpha/(1+\alpha)}\over 1 + (t/t_0)^{2\alpha/(1+\alpha)}}.
\ee
Here $t_0$ is the time at which at external observer sees 
$\Gamma_{\rm sh} = \Gamma_0$.   In the boosted frame, the shell moves forward at $t < t_0$, comes
to rest at $t = t_0$, and then begins to move backward.  The radiated Poynting flux is
concentrated around time $t_0$, where we set $x_{\rm sh} = 0$.  An analytically simple case
is $\Gamma_{\rm sh} \propto R_s^{-1}$ ($\alpha = 1$), corresponding to 
\ba
{x_{\rm sh}(t)\over ct_0} &=& 2\ln\left[{1+t/t_0\over 2}\right] - {t\over t_0} + 1\nn
\phi(t) &=& 4k_\parallel V_\parallel t_0\left[ \left({t\over t_0}\right)^{1/2} - 
\tan^{-1}\left({t\over t_0}\right)^{1/2}\right].\nn
\ea

The cumulative Poynting flux, evaluated in the boosted frame, is independent of $x$
and may therefore be obtained from 
\be
\int S_{\rm P} dt = {1\over 4\pi} \int {\partial A_z(x=0,t)\over\partial t}
                                   {\partial A_z(x=0,t)\over\partial x} dt.
\ee
The result is shown in Figure \ref{fig:poynting} for several values of $k_\parallel$ and $V_\parallel = c/\sqrt{2}$. 

\subsection{Efficiency of GHz Emission and Peak Frequency}

Each of the emission mechanisms considered here starts with a pulse of energy of width
$\sim {\cal R}$, the initial size of the dissipating field structures (colliding LSDs). 
This scale is $\sim 0.1$ cm if the collision rate is close to the observed FRB rate
(Paper II).

The low frequency electromagnetic spectrum peaks at frequency 
\be\label{eq:nup1}
\nu_{\rm pk} = {\omega_{\rm pk}\over 2\pi}
\sim {c\over 2\pi {\cal R}} = 5\times 10^{10}\,{\cal R}_{-1}^{-1}\quad{\rm Hz}.
\ee
This expression holds when magnetic drag dominates the deceleration, independent of
whether the embedded pairs gain a significant fraction of the shell energy
(whether $R_{{\rm dec},B}$ is larger than or smaller than the saturation radius
$R_{\rm sat} = 2\Gamma_{\rm max}^2{\cal R}$).   Emission at lower frequencies is less efficient,
but can easily reach $10^{39-40}$ erg if the total energy
released is ${\cal M}c^2 \sim 10^{41}{\cal M}_{20}$ erg, where ${\cal M}$ is the dipole
mass.  This leads to an interesting prediction
of relatively bright and narrow $\sim 100$ GHz pulses (Section \ref{s:hfreq}).

We consider separately the low-density (ISM-like) drag regime of Section
\ref{s:lowrho} and the high-density regime of Section \ref{s:highrho} appropriate
to the near-horizon region of a SMBH.  In the first situation, the efficiency of
GHz emission is relatively high because the ambient magnetic field is the dominant
source of drag.  In the second case, the efficiency remains high, even though
particle deflection dominates the drag,  because the ambient Alfv\'en speed is a 
sizable fraction of $c$.  

\subsubsection{Low Density -- ISM}

The shell decelerates at radius $R_{{\rm dec},B}$ given by Equation 
(\ref{eq:rdecB2}) within standard ionized ISM conditions 
($n_{\rm ex} \sim 0.01-0.1$ cm$^{-2}$ and $B_{\rm ex} \sim 1$-3 $\mu$G).
The emission frequency at radius $R_s$ is obtained from
\be
c\Delta t_{\rm em} = \int^{R_s}{d\widetilde R_s\over 2\Gamma_s^2(\widetilde R_s)} \sim {c\over \omega}.
\ee
The emission frequency inside radius $R_{{\rm dec},B}$ is
\be
\omega(R_s) \sim \omega_{\rm pk}\left({R_s\over R_{{\rm dec},B}}\right)^2\quad
(R_s < R_{{\rm dec},B}).
\ee
An asymptotic treatment of the final stages of shell deceleration outside radius 
$R_{{\rm dec},B}$ is given in Appendix \ref{a:asymp}.  Equation (\ref{eq:omega}) corresponds to
\be
\omega(R_s) \sim 2\omega_{\rm pk}\,\left({R_s\over R_{{\rm dec},B}}\right)
\left({\Gamma_s\over\Gamma_{s,\rm max}}\right)^2\quad (R_s > R_{{\rm dec},B}).
\ee
Here $\Gamma_{s,\rm max} \sim {1\over 2}(R_{{\rm dec},B}/{\cal R})^{1/2}$ is the 
peak Lorentz factor achieved by the shell, at radius $\simeq R_{{\rm dec},B}$.
The shell Lorentz factor decreases exponentially with $R_s$ beyond this point (see Equation
(\ref{eq:gamdecay})).

The energy radiated into a propagating transverse wave at the front of the shell is obtained from
\be
{d{\cal E}_{\rm rad}\over d\omega} = 
{2 R_s^2\Gamma_c^2 B_{\rm ex\perp}^2 \over |d\omega/dR_s|}.
\ee
Here $\Gamma_c$ is Lorentz factor of the front of the shell, given by Equation
(\ref{eq:gamceq}).  One finds a radiative efficiency
\be
{\omega\over {\cal E}}{d{\cal E}_{\rm rad}\over d\omega}
= K{\omega\over \omega_{\rm pk}},
\ee
with $K = 1/2$ ($1/8$) for $R_s < R_{{\rm dec},B}$ ($> R_{{\rm dec},B}$).

Finally we turn to the case where a significant fraction of the energy of the tiny
explosion is absorbed by the frozen pairs before deceleration begins.  This 
corresponds to $R_{{\rm dec},B} \gtrsim 2\Gamma_{\rm max}^2{\cal R}$, and requires
a weak ambient magnetic field, below the bound given by Equation (\ref{eq:Bmax}).
Then the Lorentz factor scales as $\Gamma_s(R_s) \simeq 
(3{\cal E}/2B_{\rm ex\perp}^2\Gamma_{\rm max}) R_s^{-3}$ in the deceleration phase.
The peak emission frequency is now obtained from
\be\label{eq:tem} 
{c\over\omega} \sim \int^R_s {d\widetilde R_s\over 2\Gamma_s(\widetilde R_s)^2} =
                 {R_s\over 14\Gamma_s^2(R_s)}.
\ee
Emission at frequency $\omega$ is concentrated in a narrow range of radius, 
$\omega \propto R_s^{-7}$.   The energy radiated is
\be
\omega {\cal E}_\omega \simeq {d\Gamma_s\over d\ln\omega} M_\pm c^2
 = {3\over 7}{\Gamma_s\over\Gamma_{\rm max}}{\cal E}.
\ee
Substituting for $R_s$ in terms of $\nu = \omega/2\pi$ and ${\cal E}$ using Equation 
(\ref{eq:tem}) gives
\be
{\omega {\cal E}_\omega\over {\cal E}}
= 0.4\,\nu_9^{3/7}{\cal E}_{41}^{1/7}\Gamma_{\rm max,7}^{-8/7}B_{\rm ex\perp,-10}^{-2/7}.
\ee
This frequency scaling was also obtained by \cite{blandford77} in the context of
black hole evaporation.

\subsubsection{High Density -- SMBH RIAF}\label{s:ghzeff}

When the ambient plasma is very dense, emission at GHz frequencies is concentrated in the final 
hydromagnetic deceleration phase.   Then the scaling (\ref{eq:gamfin})
gives
\be\label{eq:tem2}
c\Delta t_{\rm em} \sim {c\over\omega} \sim {R_s\over 8\Gamma_s^2(R_s)} 
= {\pi\rho_{\rm ex} c^2\over 6{\cal E}}R_s^4,
\ee
corresponding to $\omega \propto R_s^{-4}$. The emission radius at frequency $\omega$ is
\be\label{eq:rem}
R_{\rm em} = 5\times 10^9\,{\cal E}_{41}^{1/4} \nu_9^{-1/4} n_{\rm ex,6}^{-1/4}\quad{\rm cm}.
\ee
The emitted spectrum is, from Equation (\ref{eq:magdrag})
\be
\omega {\cal E}_\omega = {1\over 4}\cdot 2R_s^3\Gamma_s^2 B_{\rm ex\perp}^2,
\ee
and
\be\label{eq:eff}
{\omega {\cal E}_\omega\over {\cal E}} = {3B_{\rm ex\perp}^2\over 8\pi\rho_{\rm ex}c^2}.
\ee
Recent simulations of radiatively inefficient accretion flows (e.g. \citealt{yuan12}) suggest
that the magnetic energy density is a few percent of the plasma rest energy density 
near the ISCO.  

A flat radiative efficiency $\omega {\cal E}_\omega/{\cal E}$ is inferred from
Equation (\ref{eq:eff}).  This flat spectrum would extend down to the GHz band and below, 
and would extend at least to the frequency
\be
\nu(R_{\rm hydro}) \sim {c\over 2\pi} {2\Gamma_{\rm rad}^2\over R_{\rm hydro}}
= 5\times 10^{12}\;\varepsilon_{\rm th,-1}^{-4/3}{\cal E}_{41}^{1/3}
{\cal R}_{-1}^{-4/3} n_{\rm ex,6}^{1/3}\quad{\rm Hz}
\ee
emitted at the radius (\ref{eq:rdecp2}).  We infer that THz-frequency bursts could
arise from the ISCO region of slowly accreting a SMBH (paper II).

\section{Propagation Effects on the Electromagnetic Pulse}\label{s:prop}

\subsection{Damping of Electromagnetic Memory}

An interesting feature of the pulse emission mechanisms described in
Sections \ref{s:reflect} and \ref{s:em} is the net displacement
$\Delta A_\perp$ of the vector potential across the pulse (Equation
(\ref{eq:dela})).  In strong contrast with gravitational waves
\citep{christodoulou91}, this electromagnetic memory is damped by propagation through
the surrounding plasma.  

Damping of $\Delta A_\perp$ can be seen via an energetic argument as
well as directly from the wave equation (\ref{eq:wave}).  Each electron
swept up by the pulse gains a kinetic energy $(\gamma_e - 1)m_ec^2 = {1\over 2}
(\Delta a_{e\perp})^2 m_ec^2$ (Equation (\ref{eq:game2})).  This expression
holds in both the relativistic and non-relativistic regimes.  (We neglect
longitudinal polarization effects here.)
Balancing the electromagnetic pulse energy with the kinetic energy of
electrons of number density $n_{\rm ex}$ within a sphere
of radius $R_{\rm damp}$, one obtains
\ba\label{eq:rdamp}
R_{\rm damp} &=& {6c\over \omega_{P,\rm ex}^2 \Delta t_{\rm em}}\nn
&=& 5.7\times 10^{10} n_{\rm ex,0}^{-1} 
\left({\Delta t_{\rm em}\over {\rm ns}}\right)^{-1} \quad {\rm cm}.\nn
\ea
Here $\Delta t_{\rm em}$ is the duration of the pulse, which we normalize 
to the value close to the emission zone.  

Alternatively, we can integrate Equation (\ref{eq:wave}) over the width of the pulse
to obtain
\be
{1\over c}{\partial\Delta a_{e\perp}\over \partial \tau} \simeq - {\omega_{P,\rm ex}^2\over 2 c^2} 
\int_{-c\Delta t_{\rm em}}^0 a_{e\perp} d\xi \sim - {\omega_{P,\rm ex}^2\over 4c}
\Delta a_{e\perp} \Delta t_{\rm em}.
\ee
This implies a similar damping length for $\Delta a_{e\perp}$.

\subsection{Energization of Ambient Charges by Higher-Frequency Waves}\label{s:hf}

Higher-frequency components of the pulse, $\omega \gg c/\Delta R_s$, have a smaller net energizing
effect on swept-up electrons.  
The analog of Equation (\ref{eq:dela}) for a pulse of duration $\Delta t_\omega \gtrsim \omega^{-1}$ is
\be\label{eq:aom}
a_{e\perp,\omega} = {e\over m_ec\omega R_s} \left({\omega{\cal E}_\omega\over c\Delta t_\omega}\right)^{1/2}.
\ee
Before the net displacement in $a_{e\perp}$ is damped, one has $a_{e\perp,\omega} \sim 
\Delta a_{e\perp}/(\omega \Delta t_\omega)^{1/2}$.

Plasma dispersion broadens the pulse and reduces the radius at which the electron quiver motion becomes 
sub-relativistic.  The delay at frequency $\omega$ is
\be\label{eq:tdisp}
\Delta t_d \sim (1+\phi)^{-1}{\omega_{P,\rm ex}^2\over 2\omega^2}\,{R_s\over c}.
\ee
Here the electron plasma frequency of the swept-up charges is $\omega_{Pe}^2 = 4\pi n_e e^2/\gamma_e m_e  = 
4\pi n_{\rm ex} e^2/(1+\phi)m_e$.   
Given a pulse width $\Delta t_\omega \sim \Delta t_{\rm em}$ at emission, the frequency
band that maintains a geometrical overlap narrows with increasing distance from the explosion center, 
\be\label{eq:delom} 
{\Delta\omega\over\omega} \sim {\Delta t_{\rm em}\over 2\Delta t_d} 
=  {\omega^2(1+\phi)\over\omega_{P,\rm ex}^2}{c\Delta t_{\rm em}\over R_s}, 
\ee 
so that  
\be\label{eq:aom2} 
a_{e\perp,\omega} \rightarrow\left({\Delta\omega\over\omega}\right)^{1/2}a_{e\perp,\omega}. 
\ee 
The maximum relativistic quiver radius is  
\be \label{eq:rquiv}
R_{\rm rel,\omega} = \left({\omega{\cal E}_\omega\over 4\pi n_{\rm ex} m_ec^2}\right)^{1/3} = 
4.6\times10^{12}\,(\omega{\cal E}_\omega)_{39}^{1/3}n_{\rm ex,6}^{-1/3}\quad{\rm cm}. 
\ee 
This compares with 
\be R_{\rm rel,\omega} = 4\times 10^{16} \;\nu_9^{-1/2} (\omega {\cal E}_\omega)^{1/2}_{39} 
      (\omega\Delta t_{\rm em})^{-1/2} \quad {\rm cm}
\ee 
at vanishing electron density, showing that dispersion must be taken into account unless $n_{\rm ex}$ 
is much lower than in the ionized ISM.   

Higher frequency ($\sim 10^2$ GHz) emission, which we expect to dominate energetically, is less dispersed
and interacts with ambient electrons in advance of the GHz frequency component.  For example, if the LSD
size is normalized by matching the collision rate to the observed FRB rate then 
${\cal R} \sim 0.1$ cm (paper II) and the spectrum peaks at 100 GHz, with output 10-100 times stronger
than the GHz band.  This `preconditioning' of ambient electrons pushes $R_{\rm rel,\omega}$ outward by a factor
$\sim 3$-5.  

A relation between the transverse and longitudinal potentials is obtained by substituting Equation
(\ref{eq:delom}) in (\ref{eq:aom2}), 
\be\label{eq:potentials}
{a_{e\perp,\omega}^2\over 1+\phi} = {\omega{\cal E}_\omega\over 4\pi n_{\rm ex}m_ec^2 R_s^3}. 
\ee 
The net particle energy that one deduces from this expression depends on the wave amplitude.  At a
sufficiently small radius, $\phi$ saturates at $\sim m_p/2m_e$ but $a_{e\perp} \gg m_p/m_e$.  Then 
the electrons and ions reach near equipartition at energy (\ref{eq:game3}).  Given a broad-band
pulse spectrum with dispersive delay $\Delta t_d$, the energy of the particles overlapping the pulse is
\be
{{\cal E}_e + {\cal E}_p\over \omega{\cal E}_\omega} \sim 
{\langle\gamma_e\rangle\over 1+\phi}{c\Delta t_d\over R_s} \sim {c\Delta t_d\over (1-\beta_e)R_s} \ll 1.
\ee
As long as the electrons and ions are able to cross the pulse, which is the case here, they temporarily
absorb only a fraction of the wave energy.  

It should, however, be kept in mind that establishing energy equipartition between electrons and ions must involve
the excitation of some form of plasma turbulence, which may have the effect of breaking the conservation of the 
generalized transverse momentum of charged particles overlapping the pulse.  If this happens, the particles 
exiting the pulse will retain a significant fraction of the energy that they gained, with the result that
\be\label{eq:etrans}
{\cal E}_e + {\cal E}_p \sim {1\over 2}\omega {\cal E}_\omega
\ee
downstream of (inside) the pulse.

After further expansion, the wave amplitude drops into the range $(m_p/m_e)^{1/2} \lesssim a_{e\perp,\omega} 
\lesssim m_p/m_e$, while $\phi$ is still saturated near $\sim m_p/m_e$.  Now the electrostatic potential term 
in Equation (\ref{eq:game2}) for $\gamma_e$ dominates, and the energy of the overlapping particles is enhanced by a factor
$\sim (m_e a_{e\perp}/m_p)^{-2}$.  One can check that, once again, this comprises only a small fraction
of the pulse energy.   

\subsection{Rapid Transition from Relativistic to Non-Relativistic Electron Motion}\label{s:trans}

As the pulse continues to expand and approaches the relativistic quiver radius (\ref{eq:rquiv}), there is
a strong and rapid transition in collective behavior.  The pulse dispersion increases exponentially
and the quiver momentum and polarization of the swept-up particles become sub-relativistic.  To see how this happens,
note that Equation (\ref{eq:phig}) implies that $\phi$ must saturate at $\sim a_{e\perp,\omega}^2$
once $a_{e\perp,\omega}$ drops below $\sim (m_p/m_e)^{1/2}$.   This implies in turn that the right-hand side
of Equation (\ref{eq:potentials}) must be approximately unity as long as $a_{e\perp,\omega}^2$ remains large,
which clearly is possible only over a narrow range of pulse radius.   But a second solution to Equations (\ref{eq:potentials}) and (\ref{eq:phig}) is evident once the right-hand side of (\ref{eq:potentials}) drops below unity.   Then $\phi \sim a_{e\perp,\omega}^2 \sim (\omega {\cal E}_\omega)/4\pi n_{\rm ex}m_ec^2 R_s^3 \ll 1$.  

This transition is easily demonstrated by evolving the delay $\Delta t_d$ in the approximation $\phi = a_{e\perp,\omega}^2$.
Then from Equations (\ref{eq:aom}) and (\ref{eq:tdisp})
\be
{d(c\Delta t_d)\over dl} = {\omega_{p,\rm ex}^2\over 2\omega^2}\left[1 + 
         {\omega_{P,\rm ex}^2\over 2\omega^2}{R_{\rm rel,\omega}^3\over  R_s^2\,c\Delta t_d}\right]^{-1}.
\ee
Starting from the transition radius where equality of $\phi$ and $a_{e\perp,\omega}^2$ first holds, one finds
\be
{\Delta t_d\over \Delta t_{d,0}} = e^{(R_s^3-R_{s,0}^3)/3R_{\rm rel,\omega}^3}  \quad 
          (1 < \phi,\; a_{e\perp,\omega}^2 < m_p/m_e).
\ee

A more extended solution is easily obtained in the case of normal incidence of the pulse on a dense
plasma shell.  We approximate the pulse radius and electron density $n_{e,\rm sh}$ as constant in this shell, 
and once again
measure forward from the point where $\phi$ and $a_{e\perp,\omega}^2$ first reach equality.   The relativistic
transition depth is defined as $l_{\rm rel,\omega} = \omega{\cal E}_\omega/4\pi R_s^2 n_{e,\rm sh} m_ec^2$.
Then $\Delta t_d$ solves
\be
c\Delta t_d - c\Delta t_{d,0} + {\omega_{p,\rm sh}^2\over 2\omega^2} l_{\rm rel,\omega} 
          \ln\left({\Delta t_d\over \Delta t_{d,0}}\right) = {\omega_{p,\rm sh}^2\over 2\omega^2} l.
\ee
One finds exponential growth of $\Delta t_d$ at depth $l \sim l_{\rm rel,\omega}$, and linear growth thereafter.

We conclude that as the pulse expands beyond the relativistic quiver radius (\ref{eq:rquiv}),
the overlapping electrons make a rapid transition to a sub-relativistic state.  The transition
involves a drop in both the quiver amplitude and radial polarization of the pulse,
with $a_{e\perp,\omega}^2$, $\phi \propto \Delta t_d^{-1}$.   At the onset of this transition,
the electron energy is sourced mainly by the radial electrostatic potential,
$\langle\gamma_e\rangle \sim \phi/2 \sim m_p/4m_e$.  In Paper II, we show how such a rapid transition influences the
reflection of the pulse by a dense plasma cloud:  it allows the reflected pulse to avoid strong temporal
smearing and may explain the fact that the pulses emitted by the repeating FRB 121102 \citep{spitler16,scholz16} have 
durations of a few msec, and that some other FRBs show more than one component \citep{champion16}.

\subsection{Constraints on Ambient Density from \\ Synchrotron Absorption}

The accretion flow onto the Galactic Center black hole is optically thick to synchrotron absorption in
the GHz band, with an implied thermal electron density $n_e \sim 10^7$ cm$^{-3}$ near the ISCO \citep{yuan03}.
This might seem to eliminate the possibility of FRB emission near the ISCO of of a SMBH of mass 
$\sim 10^{6-8}\,M_\odot$.  However, the amplitude of the electromagnetic pulse is high enough to accelerate
ambient electrons to energies much larger than $\gamma_{\rm sync} \sim (\omega/\omega_{\rm ce})^{1/2}$,
the energy which makes the largest contribution to the synchrotron absorption coefficient.   
(Here $\omega_{\rm ce} = eB/m_ec$ is the electron cyclotron frequency.)  In this way a large-amplitude wave
can effectively `force' its way through an absorbing medium around and beyond the ISCO.

The strength of this effect, and the distance to which it extends, depends on the pulse duration and fluence, 
and the electron density in the SMBH accretion flow.
Plasma dispersion inevitably broadens the pulse and eventually forces a rapid transition to sub-relativistic
quiver motion and radial polarization (Section \ref{s:trans}).  Before this transition, which occurs around the
radius (\ref{eq:rquiv}) in a uniform medium, electrons overlapping the shell remain significantly relativistic,
$\langle\gamma_e\rangle \sim m_p/4m_e$.  We find that this energy greatly exceeds $\gamma_{\rm sync}$ at
radius $R_{\rm rel,\omega}$ when $n_{\rm ex}$ has the relatively low value characteristic of the Galactic
Center black hole, and when $M_\bullet \sim 10^6\,M_\odot$.

The accretion flow is approximated by the following radial profiles of the thermal electron density
and temperature:  $n_e = n_{e,\rm ISCO}(R/R_{\rm ISCO})^{-1}$, $\kB T_e = 10m_ec^2(R/R_{\rm ISCO})^{-1}$
(see e.g. \citealt{yuan12}).
The radius of the innermost stable circular orbit is taken for convenience to be $R_{\rm ISCO} = 6GM_\bullet/c^2$,
as appropriate for a slowly rotating black hole, but this scales out of the final results.
The energy densities of the magnetic field and the non-thermal electrons are fractions $\eta_B$,
$\eta_{\rm nth}$ of the thermal electron energy density, and the distribution function of the non-thermal
electrons is a power-law with number index $-3$.  The synchrotron absorption coefficient $\alpha_\nu$, 
taken from Equations (22) and (23) of \cite{yuan03}, is
\be
\alpha_\nu = 5.25\;\eta_{\rm nth} {e^2\, n_e \kB T_e\over m_e^2c^3\nu}\,\left({\omega_{ce}\over 2\pi\nu}\right)^3.
\ee
This implies an optical depth at a radius $r \gtrsim R_{\rm ISCO}$,
\ba\label{eq:alpsyn}
R\cdot\alpha_\nu(R) &=& 4.9\;\eta_{\rm nth} \eta_B^{3/2} {e^5 (n_e\kB T_e)^{5/2}R\over m_e^5 c^6\nu^4}\nn
          &=& 1.5\times 10^7\; {\eta_{\rm nth,-1}\,\eta_{B,-1}^{3/2}\,M_{\bullet,6}\,
              n_{e,\rm ISCO,8}^{5/2}\over \nu_9^4\,(R/R_{\rm ISCO})^4}.\nn
\ea
Here $M_{\bullet,6}$ is the SMBH mass in units of $10^6\,M_\odot$.

The electromagnetic pulse becomes weak at a distance
\be\label{eq:rrel}
{R_{\rm rel,\omega}\over R_{\rm ISCO}} = 11.5 n_{e,\rm ISCO,8}^{-1/2} M_{\bullet,6}^{-3/2} 
(\omega{\cal E}_\omega)^{1/2}_{41}
\ee
from its emission point, which also is essentially the distance from the black hole.  
Here we have normalized the pulse energy to the
expected peak of the spectrum, at $\sim 100$ GHz.  Just beyond this transition, the absorbing
electrons in the undisturbed accretion flow have an energy
\be
\gamma_{\rm sync}(R_{\rm rel,\omega}) \sim 8.6\, \eta_{B,-1}n_{e,\rm ISCO,8}^{-1/2} M_{\bullet,6}^{-3/4}
(\omega{\cal E}_\omega)_{41}^{1/4}.
\ee
This is far lower than the energy reached within the pulse just before the transition, thereby justifying our
claim that the wave is strong enough to suppress absorption.

The synchrotron optical depth is therefore dominated by particles at a distance (\ref{eq:rrel})
from the emission point.  Substituting this into Equation (\ref{eq:alpsyn}), and demanding that 
$R\cdot\alpha_\nu < 1$, gives an upper bound on the thermal electron density near the ISCO,
\be
n_{e,\rm ISCO} < 2\times 10^7\,{\nu_9^{8/9}(\omega{\cal E}_\omega)_{41}^{4/9} \over
                  M_{\bullet,6}^{14/9}\eta_{\rm nth,-1}^{2/9}\eta_{B,-1}^{1/3}}\quad {\rm cm^{-3}}.
\ee
This works out to $n_{e,\rm ISCO} \sim 10^8$ cm$^{-3}$ for a 2 GHz pulse observed from a $10^6\,M_\odot$
SMBH at a cosmological distance.

\section{Higher-Frequency Emission}\label{s:highfreq0}

Tiny explosions of the energy and size considered here will have spectral 
imprints at frequencies well above the $\sim$ GHz band in which FRBs have been detected.  
In descending order of detectability, this includes
i) higher-frequency radio-mm wave emission, ii) thermal gamma-rays from the
hot part of the explosion, and iii) synchro-curvature emission by electrons interacting
with the magnetized shell.   We now consider each of these in turn.

\subsection{Brief and Intense 0.01-1 THz Transients}\label{s:hfreq}

The electromagnetic pulse emitted during the deceleration of the shell 
has a broad spectrum (Section \ref{s:radio}).  The emission peaks at
a frequency $\nu_{\rm pk} \sim c/2\pi {\cal R} = 50\,{\cal R}_{-1}^{-1}$ GHz
in ISM-like conditions, and neglecting the effects of absorption is 
$\sim \nu_{\rm pk}/{\rm GHz})^{0-1}$ times brighter than GHz frequency emission.  
A flatter radio energy spectrum is emitted in the denser medium near a SMBH.
Since line-of-sight scattering is relatively
weak at higher frequencies, {\it the detection of such high-frequency impulses
would directly probe the size of the energy release, and therefore the energy density
(microphysical energy scale) of the underlying field structures.}   

A much higher rate of such high-frequency pulses, compared with the observed FRBs, would
also be expected if the FRB source regions are dominated by zones of high plasma density,
i.e., if many GHz pulses are eliminated from detectability by synchrotron absorption.
This is almost a necessity if most FRBs turn out to be repeating sources.  A very high
space density of LSDs would be required to produce frequent collisions, but is easily
achieved if some SMBHs form by early direct gas collapse in small galactic halos (Paper II).   If
the Galactic Center black hole had such a history, then it would be a source of
0.01-1 THz outbursts with fluences of the order of $10^{10} {\cal E}_{41} {\cal R}_{-1}$ Jy-ms!

\subsection{Prompt Fireball Gamma-rays}

The thermal component of the explosion is radiated in gamma rays of energy
$\sim 30\,\varepsilon_{\rm th,-1}^{1/4}{\cal E}_{41}^{1/4}{\cal R}_{-1}^{-3/4}$ GeV,
with an energy a fraction $\varepsilon_{\rm th}$ of the total electromagnetic energy and an essentially
unresolvable duration $\sim 3\times 10^{-12} {\cal R}_{-1}$ s.  The relatively low energy sensitivity
of gamma-ray telescopes makes this channel unpromising.  For example, on
average a detector of area $10^3$ cm$^2$ would absorb a single such gamma ray 
at a distance 1 $(\varepsilon_{\rm th,-1}{\cal E}_{41}{\cal R}_{-1})^{3/8}$ kpc from the explosion.

\subsection{Synchro-Curvature Emission from Relativistic Particles Penetrating the Shell}

Electrons swept up by the expanding, magnetized shell will radiate in response to the Lorentz force
acting on them.  The radiation energy loss is worked out in Appendix \ref{a:radloss};  it is given by
Equation (\ref{eq:dgdtrad}), which simplifies to
\be
{\partial\gamma_e\over\partial\xi}\biggr|_{\rm rad}  = {2e^2\over 3m_ec^2}
\left({\partial a_{e\perp}\over\partial\xi}\right)^2.
\ee
The corresponding frequency is given by Equation (\ref{eq:omsc}), which for a pulse of amplitude
(\ref{eq:dela}) is
\be
\hbar\omega_{s-c} = 1.2\;{\cal E}_{41}^{3/2}{\cal R}_{-1}^{1/2} R_{s,12}^{-3}\quad{\rm GeV}.
\ee
An electron that intersects the shell will suffer significant energy losses (compared with the
net kinetic energy $\sim {1\over 2}\Delta a_{e\perp}^2 m_ec^2$ acquired during transit) inside
the radius $R_{s-c}$ given by Equation (\ref{eq:rrad}),
\be
R_{s-c} = 8\times 10^{10}\;{\cal E}_{41}^{1/2}\quad{\rm cm}.
\ee

The net radiated power in synchro-curvature photons is
\be
{d{\cal E}_{s-c}\over dt} = 
4\pi n_{\rm ex} R_s^2 c \cdot \Delta R_s{d\gamma_e\over d\xi}\biggr|_{\rm rad} m_ec^2.
\ee
This works out to a very small fraction of the shell energy when the deceleration length is
larger than $R_{s-c}$, as it is in ISM-like conditions,
\be
{1\over {\cal E}}{R_s\over c}{d{\cal E}_{s-c}\over dt} = 1.1\times 10^{-3}\; {\cal E}_{41}
           {\cal R}_{-1} n_{\rm ex,0} R_{s,12}^{-1}.
\ee
On the other hand, shell deceleration in a dense medium like a SMBH accretion flow
implies a significant energy loss by the swept-up electrons (but not the ions).

\section{Ultra-High Energy Ions}\label{s:uhecr}

Ambient free ions interact strongly with the electromagnetic pulse, and effectively `surf' the pulse inside
the transit radius (\ref{eq:rtransi}).  Here we consider their dynamics, generalizing the calculation of ion 
acceleration in a persistent electromagnetic outflow by \cite{go71} to an impulsive outflow.
The maximum ion kinetic energy
is comparable to the energy that the ions gain crossing the shell for the first time at radius $R_{{\rm trans},i}$
(Equation (\ref{eq:rtransi})).
By this point the electromagnetic field is weak enough that radiation energy losses of the ions can be neglected.

We start with the radial momentum equation (Appendix \ref{a:strongem}),  
\be\label{eq:radmom}
{1\over c}{d(\gamma_i\beta_i)\over dt} = -{1\over \gamma_i}{\partial\over\partial\xi}\left({a_{i\perp}^2\over 2}\right),
\ee
where $a_{i\perp} \equiv (Zm_e/Am_p)a_{e\perp}$ and radial polarization of the shell is neglected.  We are interested
in the relativistic regime where $a_{i\perp} \gg 1$;  then 
\be\label{eq:gami}
\gamma_i \simeq (1-\beta_i^2)^{-1/2} a_{i\perp}.
\ee
For a uniformly magnetized shell,
\be
a_{i\perp} = |\xi| \, a'_{i\perp}(r) = {Ze\over Am_pc^2}\left({{\cal E}\over {\cal R}}\right)^{1/2} {|\xi|\over r}.
\ee
The displacement of an ion behind the front of the shell evolves according to
\be\label{eq:dxidt}
{1\over c}{d\xi\over dt} = -(1-\beta_i).
\ee
Substituting these approximations into Equation (\ref{eq:radmom}) and setting factors of $\beta_i \rightarrow 1$, we find
\be\label{eq:integ}
-{d\over d\ln r}\left[\xi(1-\beta_i^2)^{1/2}\right] \simeq 1.
\ee

Consider now ions that first intersect the front of the shell at radius $r_{\rm inj}$.  
Integrating Equation (\ref{eq:integ})
gives
\be\label{eq:gamrad2}
(1-\beta_i^2)^{-1/2} = {|\xi|\over \xi_{\rm inj}}{r\over r_{\rm inj}}.
\ee
The integration constant $\xi_{\rm inj}$ is obtained by considering the response of the ions near the front of the shell,
where the flow is quasi-steady.  Then $\gamma_i(1-\beta_i) \simeq 1$, which combined with $r \simeq r_{\rm inj}$ gives
\be
\xi_{\rm inj} \simeq {2\over a'_{i\perp}(r_{\rm inj})}.
\ee
Substituting this into Equation (\ref{eq:gamrad2}) and thence into Equation (\ref{eq:gami}) gives
\be
\gamma_i \simeq {1\over 2}\Bigl[\xi\,a'_{i\perp}(r_{\rm inj})\Bigr]^2.
\ee
The radial displacement $\xi$ integrates to give, 
\be
\left({|\xi|\over r_{\rm inj}}\right)^3 = {6\over [r_{\rm inj}\,a'_{i\perp}(r_{\rm inj})]^2}
\left(1 - {r_{\rm inj}\over r}\right).
\ee
The maximum displacement attained at $r \gg r_{\rm inj}$ is smaller than ${\cal R}$ inside the radius $R_{{\rm trans},i}$.  The
ion attains a maximum Lorentz factor independent of $r_{\rm inj}$,
\ba\label{eq:gamimax}
\gamma_{i,\rm max} &=& \left({9\over 2}\right)^{1/3}\Bigl[r_{\rm inj}\,a'_{i\perp}(r_{\rm inj})\Bigr]^{2/3}\nn
&=& 7.7\times 10^9\left({Z\over A}\right)^{2/3}{\cal E}_{41}^{1/3} {\cal R}_{-1}^{-1/3}.
\ea
The corresponding maximum energy is
\ba
\gamma_{i,\rm max}m_Ac^2 &=& 7.2\times 10^{18}\,{\cal E}_{41}^{1/3} {\cal R}_{-1}^{-1/3}\quad {\rm eV}\quad\quad ({\rm p})\nn
                        &=& 2.4\times 10^{20}\,{\cal E}_{41}^{1/3} {\cal R}_{-1}^{-1/3}\quad {\rm eV}\quad\quad (^{56}{\rm Fe}).\nn
\ea

The preceding results were obtained by allowing for the radial dilution of the wave amplitude, but otherwise treating
the acceleration of an embedded charge in the planar approximation.   This is valid only as long as 
$\gamma_i \lesssim r/{\cal R} = 3.7\times 10^9\,(r/R_{{\rm trans},i}) (Z/A)^{2/3} {\cal E}_{41}^{1/3} {\cal R}_{-1}^{-1/3}$.
Comparing with Equation (\ref{eq:gamimax}) shows that this condition is satisfied only when the shell has expanded
close to the maximum injection radius where the ions remain trapped within the shell.  Most of the highest-energy
ions are therefore accelerated close to radius $R_{{\rm trans},i}$.

\section{Summary}\label{s:summary}

We have described the consequences of the release of $O(10^{40-41})$ erg in electromagnetic fields within a
sub-centimeter sized volume.  A concrete proposal for achieving this is a collision of two macroscopic
magnetic dipoles (LSDs) each of mass $\sim 10^{20}$ g.  The internal magnetic field within these
relativistic field structures is $\sim 10^{20-22}$ G, some $10^{4-6}$ times stronger than any known
astrophysically (e.g. within magnetars).  

The relativistic magnetized shell produced by such a tiny explosion couples effectively to a 
low-frequency, strong, superluminal electromagnetic wave in the surrounding plasma.  
This avoids the rapid downgrading of bright radio emission by the dense plasma that is 
expected to form around bursting magnetars and colliding or collapsing neutron stars.  The closest
astrophysical emission mechanism so far proposed for FRBs invokes the coherent gyrations of ions swept up
at the front of a large ($\gtrsim$ km sized) plasmoid, e.g. following a magnetar flare \citep{lyubarsky14}.  In such
a situation, the relativistic bulk motion achieved here may not be easily repeatable, meaning that the emitted
radio pulse would be relatively broad compared with the millisecond durations of FRBs.

A summary of our main results now follows.
\vskip .05in
\noindent 1.  A small initial size ${\cal R}$ allows the outgoing relativistic plasma shell to emit
radio waves with a moderately high efficiency ($\sim ({\cal R}/\lambda)^{0-1}$), e.g. about a 1-10$\%$
efficiency for 1-10 cm waves.  The efficiency can approach unity in the 0.01-1 THz band.
\vskip .05in
\noindent 2. The measured pulse duration is a consequence of propagation effects.  In the
case of energy release within a plasma similar to the local ISM, the duration
is set by multi-path propagation through the intervening plasma.  This is consistent with
at least a subset of FRBs \citep{champion16}.  In Paper II, we consider gravitational lensing of FRB pulses
that are emitted near the ISCOs of SMBHs, as well as reflection and smearing by neighboring cold plasma.
\vskip .05in
\noindent 3. The energy release triggers the formation of a thin, magnetized, and ultrarelativistic
shell.  Ambient free electrons (ions) fully penetrate the shell outside $\sim 10^{11}$ ($10^9$) cm from the 
explosion site.  The shell experiences drag by sweeping up the ambient magnetic field, and 
by deflecting ambient electrons.  Radial polarization of the electrons and ions has an important
dynamical role in a dense medium.  In the plasma near a SMBH, the shell transitions to
a relativistic, hydromagnetic explosion outside $\sim 10^9$ cm.
\vskip .05in
\noindent 4.  We have identified three emission channels for a propagating superluminal transverse
wave outside the shell:  i) reflection of an ambient magnetic field;  ii) direct linear conversion of the
embedded magnetic field; and iii) a surface corrugation of the shell, which may be excited by
reconnection of the ejected magnetic field lines with ambient magnetic flux.  In case iii), the
superluminal mode escapes directly if the corrugation has phase speed $V_\parallel > c$, but can
also tunnel out as the shell decelerates when $V_\parallel < c$.  
\vskip .05in
\noindent 5. The spectrum and pulse shape generated through channel i) are sensitive to radial structure
in the ambient magnetic field, which can produce sharp spectral features.  Channel ii) becomes possible
if the thermal energy of the shell starts below $\sim 10^{-6}$ of the magnetic energy.  This may be a
consequence of superconductivity of the QCD vacuum near the colliding magnetic structures, in zones where
$B \sim 10^{20}$ G \citep{chernodub2010}. 
\vskip .05in
\noindent 6. The pulse is modified by transmission through surrounding plasma.  The net electromagnetic
`memory' (vector potential displacement) is damped rapidly by energy transfer to transiting electrons.
Synchrotron absorption in a plasma flow around a SMBH is negligible if the electron density at the ISCO
is below $\sim 10^7$ cm$^{-3}$ (for $M_\bullet \sim 10^6\,M_\odot$).  The rate of induced Compton 
scattering is strongly modified by the feedback of the strong wave on the transiting electrons, and
will be considered elsewhere.
\vskip .05in
\noindent 7. Primordial LSDs trace the dark matter within galaxy halos, since they interact weakly with
the ISM.  Most such broadly distributed collisions will occur within the coronal gas.  But
as is argued by in Paper II, some LSDs can be trapped within gravitationally bound cusps around
SMBHs that form by the direct collapse of massive gas clouds, producing strong collisional evolution
of the trapped LSD.  
\vskip .05in
\noindent 8. High linear polarization is a natural consequence of the model, due to the very small
emitting patch (angular size $\gamma^{-1} \sim 10^{-6}-10^{-7}$ rad).  Pulses emitted from within
the dense plasma near a SMBH will show a very high (${\rm RM} \sim 10^6$) Faraday rotation measure.
This means that sources (such as FRB 150807:  \citealt{ravi16}) with finite but small rotation measures
must be non-repeating; whereas the absence of measured polarization in the repeating FRB 121102 must
be ascribed to Faraday depolarization.
\vskip .05in
\noindent 9. Ambient ions are accelerated by surfing the expanding relativistic shell, up to $\sim 10^{19}$
eV in the case of protons and $\sim 10^{20-21}$ eV in the case of $^{56}$Fe nuclei.  
\vskip .05in
We show in a companion paper that the rate of electromagnetic pulses arriving at the Earth should be
comparable to the observed FRB rate if LSDs comprise a significant fraction of the cosmic
dark matter.  LSDs are also accreted onto massive white dwarfs ($M_{\rm wd} \gtrsim 1.0\,M_\odot$),
producing thermonuclear deflagrations or detonations at an interesting rate.

\begin{appendix}

\section{Interaction of a Strong Electromagnetic Wave with Ambient Plasma}\label{a:strongem}

Here we give a more complete description of the interaction of a strong vacuum electromagnetic pulse,
with an ambient plasma initially at rest.   We consider a simple planar wave ${\bf A}_\perp(x-ct)$, 
which is easily
generalized to a thin spherical shell.  A strong wave corresponds to $a_{e\perp} \equiv eA_\perp/m_ec^2 \gg 1$.   
We work in a gauge where $A_\perp > 0$.  
The interaction with ambient charged particles generates a radial polarization and an electrostatic potential
$\Phi$, which may also be strong in the sense that $\phi \equiv e\Phi/m_ec^2 \gg 1$.  

The wave equation in light-cone coordinates $(\xi, \tau) = (x-ct, t)$ is
\be\label{eq:wave0}
{1\over c^2}{\partial^2{\bf A}_\perp\over\partial t^2} - 
      {2\over c}{\partial^2{\bf A}_\perp\over \partial\tau\partial\xi} = {4\pi\over c}{\bf J}_\perp
   = -4\pi n_e e {\bbeta}_{e\perp}.
\ee
Conservation of canonical momentum implies that the electron quiver velocity
$\bbeta_{e\perp} = {\bf p}_{e\perp}/m_ec\gamma_e = +e{\bf A}_\perp/\gamma_em_ec^2$.  
Equation (\ref{eq:wave0}) then becomes
\be\label{eq:wave}
{\partial^2{\bf a}_{e\perp}\over\partial \tau^2} - 2c{\partial^2{\bf a}_{e\perp}\over \partial\tau\partial\xi}
   = -\omega_P^2 {{\bf a}_{e\perp}\over \gamma_e} = -\omega_{P,\rm ex}^2 {{\bf a}_{e\perp}\over 1+\phi}
\ee
Here $\omega_{P,\rm ex} = (4\pi n_{\rm ex} e^2/m_e)^{1/2}$ in the ambient cold plasma frequency.  The
second equality follows from the relation between $n_e$ and $n_{\rm ex}$ obtained in Equation
(\ref{eq:ne}) below.

The Lorentz factor of an embedded electron (charge $-e$) grows to
\be\label{eq:kin}
\gamma_e = {1\over (1 - \beta_e^2-\beta_{e\perp}^2)^{1/2}} 
= {(1+a_{e\perp}^2)^{1/2}\over(1-\beta_e)^{1/2}},
\ee
where $\beta_e$ is the speed in the direction of the wave.  The longitudinal momentum grows in
response to the non-linear Lorentz force, and is damped by radial separation of the swept-up electrons
and ions,
\be
m_e c{d(\gamma_e\beta_e)\over dt} = -e\left[E_x + (\bbeta_{e\perp}\times {\bf B}_\perp)_x\right],
\ee
where $E_x = -\partial\Phi/\partial x$ is the longitudinal electric field.  This can be rewritten as
\be
\left[{1\over c}{\partial\over\partial \tau} - (1-\beta_e){\partial\over\partial\xi}\right](\gamma_e\beta_e)
= {\partial\phi\over\partial\xi} - {1\over\gamma_e}{\partial\over\partial\xi}\left({a_{e\perp}^2\over 2}\right).
\ee
In light-cone coordinates,
\be\label{eq:euler}
{\partial\over\partial\xi}\left[\gamma_e(1-\beta_e)-\phi\right] 
= -{1\over c}{\partial(\gamma_e\beta_e)\over\partial\tau}.
\ee
The continuity equation of the swept-up electrons similarly can be written
\be\label{eq:cont}
{\partial\over\partial\xi}\left[(1-\beta_e)n_e\right] = {1\over c}{\partial n_e\over\partial\tau}.
\ee

A simple steady solution for $\gamma_e$, $\beta_e$, and $\phi$ is easily found in this cartesian approximation,
corresponding in a spherical geometry to a short transit time of the electrons across the pulse compared with 
the expansion time $\tau \sim r/c$.  Letting the right-hand sides of Equation (\ref{eq:euler}) vanish, we find
the constraint $\gamma_e(1-\beta_e) = 1+\phi$, which is combined with Equation (\ref{eq:kin}) to give
\be\label{eq:game2}
\gamma_e = {1+a_{e\perp}^2\over 2(1+\phi)} + {1+\phi\over 2}.
\ee
Similarly 
\be\label{eq:ne}
n_e = {n_{\rm ex}\over 1-\beta_e} = {\gamma_e\over 1+\phi}n_{\rm ex}.
\ee
The dynamics of the ions is computed similarly, with the result (here $\mu \equiv Zm_e/Am_p$)
\be
\gamma_i = {1+\mu^2a_{e\perp}^2\over 2(1-\mu\phi)} + {1-\mu\phi\over 2}
\ee
and 
\be
Zn_i = {\gamma_i\over 1-\mu\phi}n_{\rm ex}.
\ee

The electrostatic potential solves
\be
{\partial^2\Phi\over\partial\xi^2} = 4\pi n_{\rm ex}e(n_e-Zn_i),
\ee
or equivalently
\be
{\partial^2\phi\over\partial\xi^2} = {\omega_{P,\rm ex}^2\over 2c^2}\left[{1+a_{e\perp}^2\over (1+\phi)^2} - 
      {1 + \mu^2 a_{e\perp}^2\over (1-\mu\phi)^2}\right].
\ee
This equation may be integrated to give
\be\label{eq:phig}
\left({\partial\phi\over\partial\xi}\right)^2 = {\omega_{P,\rm ex}^2\over c^2}
     \phi\left[{1+a_{e\perp}^2\over 1+\phi} - {1 + \mu^2a_{e\perp}^2\over 1-\mu\phi}\right].
\ee

There are two sets of circumstances where the longitudinal potential grows to $\phi > 1$.
First, the ambient electron density is low and dispersion remains too weak to significantly
broaden the pulse beyond $\Delta t_\omega \sim \omega^{-1}$ while its amplitude remains high.  
Then radial polarization is important only close to the emission site, where $a_{e\perp} > 
\omega/\omega_{P,\rm ex}$.  One has $\phi \simeq a_{e\perp}\omega_{P,\rm ex}\Delta t_\omega$
up to a maximum $\phi \sim \mu^{-1}$ where the motion of the ions and electrons become coupled.  

The singularity in Equation (\ref{eq:phig}) limits the further growth of $\phi$.  
Since $\mu \ll 1$, the numerical solution to this equation shows nearly linear sawtooth behavior
when the gradient scale of $\phi$ is small compared with that of $a_{e\perp}$.  Then the radial
speeds of the electrons and ions equilibrate at
\ba\label{eq:betaeq}
\langle 1-\beta_e\rangle &=& {2\langle\phi^2\rangle\over a_{e\perp}^2} = {2\over 3}\left({Am_p\over Zm_e a_{e\perp}}\right)^2;\nn
\langle 1-\beta_i\rangle  &=& {2\langle(1-\mu\phi)^2\rangle\over \mu^2a_{e\perp}^2} = \langle 1-\beta_e\rangle.
\ea
Taking this as the speed of a radially equilibrated flow, and substituting a mean potential $\phi = 1/2\mu$,
the particle energies average to
\be\label{eq:game3}
\langle\gamma_e\rangle = {3\mu\over 4} a_{e\perp}^2 = {1\over \mu}\langle\gamma_i\rangle,
\ee
implying energy equipartition between electrons and protons.

A second pattern of collective behavior is found when the ambient electron density is high, so that the
pulse energy is stretched significantly by plasma dispersion.  The preceding solution is recovered
when $a_{e\perp} > \mu^{-1}$.  The longitudinal potential still reaches the limiting value $\mu^{-1}$
in the intermediate regime $\mu^{-1/2} < a_{e\perp} < \mu^{-1}$, but now the longitudinal piece of
the electron energy (\ref{eq:game2}) dominates, One finds instead
\be 
\langle\gamma_e\rangle \sim {1\over 4\mu};  \quad  \langle\gamma_i - 1\rangle \sim {1\over 4}.
\ee
When $a_{e\perp} < \mu^{-1/2}$, inspection of equation (\ref{eq:phig}) shows that $\phi < a_{e\perp}^2 
< \mu^{-1}$.  We argue in Section \ref{s:hf} that a pulse amplitude $1 < a_{e\perp} < \mu^{-1/2}$
cannot be sustained when the pulse is strongly dispersed:  the wave is unstable to shifting to 
a much lower polarization and amplitude, $\phi, a_{e\perp} < 1$.

\subsection{Distinction between Relativistic Magnetized Shell and Superluminal Electromagnetic Pulse}

The formulae just presented do not depend explicitly on the density of electrons and positrons carried
outward with the electromagnetic pulse.  They therefore apply to both the initial MHD impulse, and to a
superluminal electromagnetic mode excited by the interaction with ambient plasma.  The initial impulse
is very narrow, $\Delta R_s \sim 0.01-0.1$ cm, and the electromagnetic response is somewhat broader,
with duration $\Delta t_{\rm em} \sim \omega^{-1}$ at frequency $\omega$.  The relative wave amplitudes are
\be
{a_{e\perp,\omega}\over \Delta a_{e\perp}}  \sim  \left({\omega{\cal E}_\omega\over {\cal E}}\right)^{1/2}
\left({\omega {\cal R}\over c}\right)^{-1/2}.
\ee

\subsection{Radiative Losses of the Embedded Electrons}\label{a:radloss}

The electrons swept up by the shell are strongly accelerated and must radiate part or most of their energy.
This loss is calculated using the relativistic Larmor formula, expressed in terms of the four-velocity 
$u_e^\mu = \gamma_e(1,\beta_e, \bbeta_{e\perp})$, 
\be
{d\gamma_e\over dt}\biggr|_{\rm rad} \simeq -(1-\beta_e)c{\partial\gamma_e\over\partial\xi}\biggr|_{\rm rad}
= -{2e^2\over 3m_ec^3}\gamma_e^2{du_e^\mu\over dt}{du_{e\,\mu}\over dt}.
\ee
Here we make the thin-shell approximation, taking $\partial/\partial\tau \rightarrow 0$ in light-cone
coordinates.  The four-acceleration $du_e^\mu/dt$ is calculated from a non-radiating particle trajectory.
Hence we substitute
\be
{d{\bf u}_{e\perp}\over dt} = {d{\bf a}_{e\perp}\over dt}; \quad 
{1\over c}{d(\gamma_e\beta_e)\over dt} = {\partial\phi\over\partial\xi} - 
{1\over 2\gamma_e}{\partial a_{e\perp}^2\over\partial\xi};  \quad
{d\gamma_e\over dt} = {1\over 2\gamma_e}{d\over dt}\left[u_{e\perp}^2 + (\gamma_e\beta_e)^2\right]
\ee
to get
\be\label{eq:dgdtrad}
{\partial\gamma_e\over\partial\xi}\biggr|_{\rm rad} = {2e^2\over 3m_ec^2}
\left[\gamma_e^2(1-\beta_e) + a_{e\perp}^2-u_{e\perp}^2 + \beta_e(a_{e\perp}-u_{e\perp})^2\right]
     \left({\partial a_{e\perp}\over\partial \xi}\right)^2.
\ee
We have further assumed that the wave field $a_{e\perp}$ varies on a short lengthscale $c/\omega$ 
compared with the gradient length of the electrostatic potential, allowing us to take $\partial\phi/\partial\xi
\rightarrow 0$ in the Larmor formula.

The transverse momentum is only slightly perturbed when the radiation loss is a perturbation to the
particle trajectory.  We therefore substitute $u_{e\perp} \simeq a_{e\perp}$ in Equation (\ref{eq:dgdtrad}),
and use $d(\gamma_e\beta_e)/dt|_{\rm rad} = \beta_e d\gamma_e/dt|_{\rm rad}$ to get
\be
{\partial\over\partial\xi}\left[\gamma_e(1-\beta_e)\right] \simeq {\partial\phi\over\partial\xi}
 + {1\over \xi_{\rm rad}}\left[\gamma_e(1-\beta_e)\right]^2.
\ee
Recall that the particles swept up by the electromagnetic pulse flow toward negative $\xi$,
meaning that radiation damping generates a positive $\xi$-gradient in particle energy.
In a spherical geometry, the radiation damping length
\be
\xi_{\rm rad}^{-1}= {2e^2\over 3m_ec^2}\left({\partial a_{e\perp}\over\partial\xi}\right)^2
\ee
can be easily expressed in terms of the radially integrated compactness $\ell$,
\be\label{eq:rrad}
{\xi_{\rm rad}\over\Delta R_s} = \ell^{-1} \equiv \left({\sigma_T {\cal E}\over 4\pi r^2 m_ec^2}\right)^{-1}
     = {r^2\over R_{s-c}^2}; \quad\quad   R_{s-c} = 8\times 10^{10}\;{\cal E}_{41}^{1/2}\quad{\rm cm}.
\ee
This expression follows directly from Equation (\ref{eq:dela}) in the case of a uniformly magnetized
shell, but also applies to a vacuum electromagnetic pulse composed of modes of frequency $\gg c/\Delta R_s$.

The frequency of the radiated photons is next obtained by noting that the transverse force
$du_{e\perp}/dt$ makes the dominant contribution to the Larmor power.  In this situation, the 
curvature frequency\footnote{Equivalent to $c/R_c$ in the case of electron motion on a fixed
trajectory with radius of curvature $R_c$.} $\omega_c$ is related to the emitted power by
\be
m_ec^2{d\gamma_e\over dt}\biggr|_{\rm rad} = -{2e^2\over 3c} \gamma_e^4\omega_c^2,
\ee
and the synchro-curvature frequency is
\be\label{eq:omsc}
\omega_{s-c} \sim 0.3\gamma_e^3\omega_c = 0.3\cdot {c\over 2}a_{e\perp}^2
\left|{\partial a_{e\perp}\over\partial\xi}\right|.
\ee
This last identity makes use of Equation (\ref{eq:game}).

\section{Electromagnetic Emission During Final Shell Deceleration: ISM Conditions}\label{a:asymp}

Here we derive an analytic approximation to the outgoing electromagnetic spectrum produced,
in ISM-like conditions, by the relativistic reflection of the ambient magnetic field.   The initial
deceleration length is small enough ($R_{{\rm dec},B} \ll 2\Gamma_{\rm max}^2{\cal R}$) that
one must take into account the reduction in Lorentz factor in a forward part of the relativistic shell 
due to its interaction with the ambient magnetic field.  We treated this interaction in Section \ref{s:diffaccel}
by balancing the momentum flux on either side of the contact discontinuity separating the relativistic
shell from the reflected ambient magnetic field.   The resulting deceleration length is given by
Equation (\ref{eq:rdecB2}).  Near this radius, the shell Lorentz factor reaches a maximum
\be
\Gamma_{s,\rm max} \simeq {1\over 2}\left({R_{{\rm dec},B}\over {\cal R}}\right)^{1/2}.
\ee

We now follow the deceleration to its final stages at $R_s > R_{{\rm dec},B}$, taking into account the transition to
electron-dominated drag.  This means that we must now balance all three terms on the right-hand side of Equation 
(\ref{eq:netforcec}).  In this late stage, we take the Lorentz factor $\Gamma_s$ to be uniform within the shell,
and assume that its magnetization remains high.  (The numerical solution shown in Figures \ref{fig:shellkin}
and \ref{fig:shellkin2} relaxes the latter assumption.)  We solve first for the rest-frame shell thickness 
$\Delta R_s' = \Gamma_s \Delta R_s$, and then for the Lorentz factor, which is obtained from
\be\label{eq:gamsdyn}
\Gamma_s^{-1} \simeq {d\Delta R_s'\over dR_s}.
\ee
Equation (\ref{eq:netforce}) can be written as
\be
{1\over \Gamma_s^2 (\Delta R_s')^2}\left[1-{8\pi e^2 n_{\rm ex}\over m_ec^2}(\Delta R_s')^2\right] = 
     4{R_s^2\over R_{{\rm dec},B}^2}.
\ee
The electron drag, represented by the term proportional to $n_{\rm ex}$, implies a maximum comoving
shell thickness $\Delta R_{s,\rm max}' = (m_ec^2/8\pi e^2 n_{\rm ex})^{1/2}$, near which 
$\Gamma_s \rightarrow 0$ and the shell dissipates ($\Delta R_s \rightarrow \infty$).  
Combining the above two equations gives
\be\label{eq:eom2}
\left[1 - {(\Delta R_s')^2\over (\Delta R_{s,\rm max}')^2}\right]^{1/2} {d\ln\Delta R_s'\over dR_s} = 
   {2R_s\over R_{{\rm dec},B}^2}.
\ee
This equation is readily integrated, but we will be satisfied with two limiting solutions.

As long as $\Delta R_s' \ll \Delta R_{s,\rm max}'$, one finds
\be
{\Delta R_s'\over \Gamma_{s,\rm max} {\cal R}} \simeq \exp\left({R_s^2\over R_{{\rm dec},B}^2} - 1\right)
\ee 
and, from Equation (\ref{eq:gamsdyn}),
\be\label{eq:gamdecay}
{\Gamma_s\over\Gamma_{s,\rm max}} \simeq {2R_s\over R_{{\rm dec},B}}\, \exp\left(1-{R_s^2\over R_{{\rm dec},B}^2}\right).
\ee
The frequency of the radiation emitted at radius $R_s$ and Lorentz factor $\Gamma_s$ is obtained from
\be\label{eq:omega}
{c\over\omega} \sim \int^{R_s} {d\widetilde R_s\over 2\Gamma_s^2(\widetilde R_s)} \simeq {{\cal R}\over 2} 
    \left({R_s\over R_{{\rm dec},B}}\right)^{-1} \left({\Gamma_s\over \Gamma_{s,\rm max}}\right)^{-2}.
\ee

The comoving shell thickness approaches its limiting value $\Delta R_{s,\rm max}'$ at a radius
\be
R_{{\rm dec},e}  =  
R_{{\rm dec},B} \left[\ln\left(2{\Delta R_{s,\rm max}'\over \Gamma_{s,\rm max} {\cal R}}\right)\right]^{1/2},
\ee
close to which
\be
{\Delta R_s'\over \Gamma_{s,\rm max} {\cal R}} \simeq {1\over 2} \exp\left({R_s^2\over R_{{\rm dec},B}^2}\right).
\ee
The Lorentz factor is, from Equations (\ref{eq:gamsdyn}) and (\ref{eq:eom2}),
\be
\Gamma_s(R_s)  = {R_{{\rm dec},B}^2\over 2 R_s \Delta R'_s(R_s)} 
           \left\{1 - {[\Delta R_s'(R_s)]^2\over (\Delta R'_{s,\rm max})^2}\right\}^{1/2}.
\ee

\end{appendix}

\end{document}